\documentclass[a4paper,fleqn]{cas-sc}

\usepackage[numbers,sort&compress]{natbib}
\usepackage{amsmath,amssymb,bm}
\usepackage{mathtools}
\usepackage{algpseudocode}

\newcommand{\B}{\bm{B}}
\newcommand{\dij}{\bm{d}_{ij}}
\newcommand{\dji}{\bm{d}_{ji}}
\newcommand{\ri}{\bm{r}_i}
\newcommand{\rj}{\bm{r}_j}
\newcommand{\G}{\mathcal{G}}

\newcommand{\Np}{N_{\rm p}}
\newcommand{\Bnp}{\B_{[\Np]}}
\newcommand{\dd}{\mathrm{d}}

\begin{document}
\let\WriteBookmarks\relax
\def\floatpagepagefraction{1}
\def\textpagefraction{.001}

\shorttitle{Adjoint Projection for SPMHD}
\shortauthors{Y. Tsukamoto}

\title[mode=title]{An Adjoint Projection Formulation for Enforcing the\texorpdfstring{\\}{ }
\texorpdfstring{\(\nabla\cdot\mathbf{B}=0\)}{divergence-free} Constraint in Smoothed Particle Magnetohydrodynamics}

\author[1]{Yusuke Tsukamoto}
\cormark[1]
\ead{tsukamoto@konan-u.ac.jp}
\affiliation[1]{organization={Faculty of Science and Engineering, Konan University},
            addressline={Okamoto 8-9-1, Higashinada-ku},
            city={Kobe},
            postcode={658-8501},
            country={Japan}}
\cortext[cor1]{Corresponding author}

\begin{abstract}
We present a projection method for controlling numerical \(\nabla\cdot\B\) errors in smoothed particle magnetohydrodynamics (SPMHD).
The method corrects the magnetic field after an MHD update by solving an elliptic projection problem constructed from the same discrete divergence operator used to measure the error.
A key ingredient is to use the adjoint gradient associated with a volume-weighted metric.
With this choice, the projection gives an energy-minimizing correction, does not increase the discrete magnetic energy, and leads to a symmetric positive semidefinite linear system that can be solved by the conjugate-gradient method without explicitly assembling the matrix.
We test the method using two-dimensional Dedner-type divergence tests and three-dimensional magnetized collapse calculations.
With sufficiently many iterations, the projection reduces the divergence error to the floating-point roundoff level in both test problems.
In realistic collapse runs, practical stopping criteria designed to reduce the divergence error generated by the underlying SPMHD update suppress the normalized divergence error well below that obtained in the divergence-cleaning run, with a projection cost of only about \(1\)--\(10\%\) of the SPMHD update cost.
The density and plasma-\(\beta\) structures remain consistent when the projection interval is varied, whereas the divergence-cleaning run shows quantitative differences.
These results indicate that the projection method is a robust and attractive alternative to divergence cleaning for controlling \(\nabla\cdot\B\) errors in SPMHD and related particle or meshless MHD schemes.
\end{abstract}

\begin{keywords}
magnetohydrodynamics (MHD) \sep numerical methods \sep smoothed particle hydrodynamics \sep divergence cleaning \sep projection methods
\end{keywords}

\maketitle

\section{Introduction}

The solenoidal constraint on the magnetic field,
\begin{equation}
    \nabla\cdot\B = 0,
\end{equation}
is a central numerical issue in magnetohydrodynamics (MHD).
In mesh-based MHD codes, constrained transport (CT) has been the most promising and widely used approach because it preserves the discrete divergence through the topological pairing of face-centered magnetic fluxes and edge-centered electromotive forces \citep{1988ApJ...332..659E}.

In meshless methods, however, this CT construction is not straightforward to apply because particles do not define clear control-volume boundaries or a natural face--edge topology on which magnetic fluxes and electromotive forces can be evolved.
Divergence cleaning, which evolves a cell- or particle-centered auxiliary scalar field through hyperbolic propagation and parabolic damping of divergence errors \citep{2002JCoPh.175..645D}, is therefore the mainstream approach in smoothed particle magnetohydrodynamics (SPMHD) and closely related meshless MHD schemes.
Modern SPMHD commonly uses constrained hyperbolic/parabolic cleaning \citep{2005MNRAS.364..384P,2012JCoPh.231..759P,2012JCoPh.231.7214T,2016JCoPh.322..326T,2018PASA...35...31P}, and Godunov-type SPMHD (GSPMHD) also uses Dedner-type hyperbolic divergence cleaning \citep{2011MNRAS.418.1668I,2013ASPC..474..239I}.
Dedner-type cleaning, however, contains two parameters with physical dimensions: the propagation speed of the auxiliary cleaning field and the damping length scale, or equivalently the damping timescale associated with that length.
These parameters determine how rapidly divergence errors are transported and damped.
In simulations with a large dynamical range, the appropriate values can depend on position and on the evolutionary stage of the system.
The original Dedner formulation assumes these parameters to be spatially and temporally constant; if they are varied without a consistent treatment, their variation can instead become a source term for \(\nabla\cdot\B\).
The choice of cleaning speed and damping timescale can affect astrophysical MHD calculations, as recently pointed out by \citet{2026arXiv260507928T}.

Projection methods instead solve an elliptic problem after a time update and map the magnetic field back to a discrete divergence-free subspace.
Such projection ideas are well established in grid-based MHD divergence control \citep[e.g.][]{2000JCoPh.161..605T}.
In SPMHD, \citet{2005MNRAS.364..384P} tested a projection approach based on a gravity-like direct \(O(N^2)\) summation discretization for the scalar correction.
They found that the method was reasonably effective for divergence errors with wavelengths larger than the smoothing length, but was less effective for short-wavelength, \(\sim h\), noise because the SPH operator used to estimate \(\nabla\cdot\B\) smooths such noise.
This behavior may be attributed to a discretization-level inconsistency between the divergence operator used to measure the error and the gravity-like operator used to compute the scalar correction.
To our knowledge, however, an SPMHD projection formulation that keeps divergence and gradient operators mutually consistent by constructing the gradient correction as the adjoint of the chosen discrete \(\nabla\cdot\B\) operator has not yet been examined.
This gap motivates the projection method developed in this paper.

This paper is organized as follows.
Section 2 presents the formulation and discretization of the projection method: Section 2.1 formulates the residual projection problem, Section 2.2 introduces the SPH divergence operator and its volume-metric adjoint gradient, Section 2.3 summarizes the mathematical properties of the projection, and Section 2.4 describes the numerical implementation and pseudocode.
Section 3 presents numerical tests, and Section 4 summarizes the results and discusses their implications.

\section{Formulation and Discretization}

\subsection{Projection as Residual Reduction}

\subsubsection{Unprojected magnetic field}

Let \(\B(\bm{x},t)\) denote the continuum magnetic field, and let \(\B_i^n\) be its value carried by particle \(i\) at time \(t_n\).
To distinguish a continuum field from the finite-dimensional particle vector, we write the stacked particle field as
\begin{equation}
    \Bnp^n
    =
    (\B_1^n,\B_2^n,\ldots,\B_{\Np}^n)^T
    \in\mathbb{R}^{3\Np}.
    \label{eq:stacked_B_def}
\end{equation}
The bracketed subscript \([\Np]\) labels the number of particles in the stacked vector and is not a particle index.
We first apply an arbitrary induction update before projection and define the resulting unprojected particle vector as
\begin{equation}
    \Bnp^* = T_B^n \Bnp^n.
    \label{eq:unprojected_update}
\end{equation}
Here \(T_B^n\) denotes the induction operator used by the underlying SPMHD code.
It may correspond to the induction update of standard SPMHD \citep{2005MNRAS.364..384P,2012JCoPh.231..759P} or Godunov-type SPMHD \citep{2011MNRAS.418.1668I,2013ASPC..474..239I}, with or without additional dissipative terms.
The projection method described below does not require specifying \(T_B^n\), except through the unprojected particle values in \(\Bnp^*\).

Let \(D\) be the discrete SPH divergence operator selected for the solenoidal constraint.
For a fixed particle distribution and fixed SPH derivative coefficients during one projection step, \(D\) is a linear map \(D:\mathbb R^{3N_{\rm p}}\rightarrow\mathbb R^{N_{\rm p}}\) from the stacked particle vector field to the particle scalar field.
Its explicit SPH form is given below.
The unprojected divergence residual is the \(N_{\rm p}\)-dimensional vector \(r\in\mathbb R^{N_{\rm p}}\) defined by
\begin{equation}
    r = D\Bnp^*,\qquad r_i = (D\Bnp^*)_i.
    \label{eq:unprojected_residual}
\end{equation}
Equations involving \(D\), \(\G\), \(\pi\), and \(r\) below are written in this stacked-vector notation unless otherwise stated.
The goal of the projection is to reduce this residual to zero, or to the tolerance of the elliptic solver.

\subsubsection{Residual correction and the roles of \texorpdfstring{\(\G\)}{G} and \texorpdfstring{\(\pi\)}{pi}}

We seek the corrected field in the form
\begin{equation}
    \Bnp^{n+1}=\Bnp^* - \G\pi.
    \label{eq:projection_ansatz}
\end{equation}
Here \(\pi\) is the scalar Lagrange-multiplier field associated with the constraint \(D\Bnp=0\).
The stacked vector \(\G\pi\in\mathbb R^{3N_{\rm p}}\) is the projection correction generated from this scalar multiplier.
The operator \(\G:\mathbb R^{N_{\rm p}}\rightarrow\mathbb R^{3N_{\rm p}}\) will be defined as the adjoint gradient associated with the selected divergence operator \(D\).
It should not be replaced by an arbitrary SPH gradient.

The constraint is
\begin{equation}
    D\Bnp^{n+1}=0.
\end{equation}
Substituting equation (\ref{eq:projection_ansatz}) gives
\begin{equation}
    D(\Bnp^* - \G\pi)=0,
\end{equation}
or
\begin{equation}
    D\G\pi = D\Bnp^* = r.
    \label{eq:projection_equation_abstract}
\end{equation}
Thus the projection step is the solution of the residual equation
\begin{equation}
    L\pi=r, \qquad L\equiv D\G.
\end{equation}
In practice, an iterative method constructs
\begin{equation}
    \Bnp^{(m)}=\Bnp^*-\G\pi^{(m)}
\end{equation}
and reduces \(D\Bnp^{(m)}\) until it is sufficiently small.

\subsubsection{Constrained minimization and the volume metric}
\label{subsec:constrained_minimization}

The constraint \(D\Bnp^{n+1}=0\) alone does not, in general, uniquely determine the projected magnetic field.
This is because it imposes at most \(N_{\rm p}\) scalar constraints on the \(3N_{\rm p}\) degrees of freedom of the stacked magnetic-field vector.
We therefore regard the projection as a constrained minimization problem that minimizes the change in the discrete magnetic-energy norm under the constraint \(D\Bnp^{n+1}=0\).
Through this minimization problem, we determine the appropriate form of the correction operator \(\G\).
Using the stacked notation introduced in equation~(\ref{eq:stacked_B_def}), this minimization problem is written as
\begin{equation}
    \Bnp^{n+1}
    =
    \arg\min_{\widetilde{\B}_{[\Np]}}
    \frac{1}{2}\sum_i V_i |\widetilde{\B}_i-\B_i^*|^2
    \quad
    \text{subject to}
    \quad
    D\widetilde{\B}_{[\Np]}=0.
    \label{eq:argmin_Bi}
\end{equation}
Here the argument of the minimization is the entire stacked trial vector \(\widetilde{\B}_{[\Np]}\).
The result is the stacked vector \(\Bnp^{n+1}\), whose particle-wise components \(\B_i^{n+1}\) give the minimizing magnetic field over all particles and satisfy \(D\Bnp^{n+1}=0\).
We use
\begin{equation}
    V_i=\frac{m_i}{\rho_i}
    \label{eq:volume_def}
\end{equation}
as the particle volume.
Equations~(\ref{eq:argmin_Bi}) and (\ref{eq:volume_def}) therefore define the discrete inner product, or metric, used by the projection:
\[
    \langle X,Y\rangle_V \equiv \sum_i V_i \bm{X}_i\cdot\bm{Y}_i,
    \qquad
    \|X\|_V^2 \equiv \langle X,X\rangle_V .
\]
The objective in equation~(\ref{eq:argmin_Bi}) is \( \frac{1}{2}\|\widetilde{\B}_{[\Np]}-\Bnp^*\|_V^2 \).
This is an essential part of the method because it fixes the metric of the projection.
With this choice,
\begin{equation}
    \sum_i V_i |\Delta\B_i|^2 \approx \int |\Delta\B(\bm{x})|^2\,\dd V,
\end{equation}
where \(\Delta\B(\bm{x})\) is the corresponding correction field in the continuum description and \(\Delta\B_i\equiv \B_i^{n+1}-\B_i^*\) are its particle values.
Since the magnetic energy is
\begin{equation}
    E_B = \frac{1}{8\pi}\int |\B(\bm{x})|^2\,\dd V,
\end{equation}
the volume metric is the natural metric for minimizing the magnetic-field correction.
By contrast, a mass metric, \(\sum_i m_i|\Delta\B_i|^2\), would approximate \(\int \rho |\Delta\B(\bm{x})|^2\,\dd V\), i.e. a density-weighted correction norm.
This may be natural for velocity projection, but it is not the magnetic-energy norm.

Let
\begin{equation}
    \Delta\B_{[\Np]}=\Bnp^{n+1}-\Bnp^*.
\end{equation}
The constraint becomes
\begin{equation}
    D(\Bnp^*+\Delta\B_{[\Np]})=0,
\end{equation}
or
\begin{equation}
    D\Delta\B_{[\Np]}=-D\Bnp^*=-r.
\end{equation}
We introduce a scalar Lagrange multiplier \(\pi_i\) and define the Lagrangian
\begin{equation}
    \mathcal{L}
    =
    \frac{1}{2}\sum_i V_i |\Delta\B_i|^2
    +
    \sum_i \pi_i\left[(D\Delta\B_{[\Np]})_i+r_i\right].
    \label{eq:Lagrangian_discrete}
\end{equation}
In matrix notation this is
\begin{equation}
    \mathcal{L}
    =
    \frac{1}{2}\Delta\B_{[\Np]}^T M_V \Delta\B_{[\Np]}
    +
    \pi^T(D\Delta\B_{[\Np]}+r),
    \label{eq:Lagrangian_matrix}
\end{equation}
where the vector-field mass matrix is
\begin{equation}
    M_V={\rm diag}(V_1 I_3,V_2 I_3,\ldots,V_{N_{\rm p}}I_3).
    \label{eq:MV_def}
\end{equation}
Here \(I_3\) is the \(3\times 3\) identity matrix.
Explicitly,
\begin{equation}
M_V =
\begin{pmatrix}
V_1 I_3 & 0 & 0 & \cdots \\
0 & V_2 I_3 & 0 & \cdots \\
0 & 0 & V_3 I_3 & \cdots \\
\vdots & \vdots & \vdots & \ddots
\end{pmatrix},
\end{equation}
or, in components,
\begin{equation}
    (M_V)_{(i,\alpha),(k,\beta)}=V_i\delta_{ik}\delta_{\alpha\beta}.
\end{equation}
Here \(i\) and \(k\) label particles, while \(\alpha\) and \(\beta\) label Cartesian components.
Varying the stacked vector \(\Delta\B_{[\Np]}\) gives, using \(M_V^T=M_V\),
\begin{equation}
    \delta\mathcal{L}
    =
    \left(M_V\Delta\B_{[\Np]} + D^T\pi\right)^T
    \delta\Delta\B_{[\Np]}.
\end{equation}
Because \(\delta\Delta\B_{[\Np]}\) is arbitrary, stationarity gives
\begin{equation}
    M_V\Delta\B_{[\Np]} + D^T\pi
    =
    0.
\end{equation}
Thus
\begin{equation}
    \Delta\B_{[\Np]} = -M_V^{-1}D^T\pi.
\end{equation}
This motivates the definition
\begin{equation}
    \G \equiv M_V^{-1}D^T.
    \label{eq:G_def}
\end{equation}
This identifies the abstract adjoint form of the correction operator.
Then
\begin{equation}
    \Delta\B_{[\Np]}=-\G\pi,
\end{equation}
and therefore
\begin{equation}
    \Bnp^{n+1}=\Bnp^* - \G\pi.
\end{equation}
Substituting this into the divergence constraint gives
\begin{equation}
    D\G\pi=D\Bnp^*=r.
    \label{eq:projection_equation}
\end{equation}
This is the projection equation.

\subsection{SPH Divergence and Adjoint Gradient}

The purpose of this subsection is to construct the explicit SPH form of the correction operator \(\G\) associated with the chosen divergence operator \(D\).

\subsubsection{Discrete divergence operator}

The neighbor list of particle \(i\) is denoted by \(\mathcal N_i\).
We assume that the neighbor graph is symmetric,
\begin{equation}
    j\in\mathcal N_i \Longleftrightarrow i\in\mathcal N_j.
\end{equation}
The discrete operator acting on a stacked particle vector \(\Bnp\) is defined by
\begin{equation}
    (D\Bnp)_i
    =
    \sum_{j\in\mathcal N_i}
    \frac{m_j}{\Omega_i\rho_i}
    \nabla_i W(|\ri-\rj|,h_i)
    \cdot(\B_j-\B_i)
    \equiv
    \sum_{j\in\mathcal N_i}
    \dij\cdot(\B_j-\B_i),
    \label{eq:D_action}
\end{equation}
where
\begin{equation}
    \dij
    =
    \frac{m_j}{\Omega_i\rho_i}
    \nabla_i W(|\ri-\rj|,h_i).
    \label{eq:dij_def}
\end{equation}
The reverse directed coefficient, obtained by exchanging \(i\) and \(j\), is
\begin{equation}
    \dji
    =
    \frac{m_i}{\Omega_j\rho_j}
    \nabla_j W(|\rj-\ri|,h_j).
    \label{eq:dji_def}
\end{equation}
In general,
\begin{equation}
    \dij \neq \dji,
\end{equation}
because the derivative coefficients are evaluated with particle-dependent smoothing lengths, densities, and \(\Omega\)-factors.

\subsubsection{Component and matrix form}

In components,
\begin{equation}
    (D\Bnp)_i
    =
    \sum_{j\in\mathcal N_i}
    \sum_{\alpha\in\{x,y,z\}}
    d_{ij}^{\alpha}(B_j^{\alpha}-B_i^{\alpha}),
    \label{eq:D_action_components}
\end{equation}
where the Cartesian component index \(\alpha\) runs over \(x,y,z\).

As a matrix,
\begin{equation}
    D\in\mathbb R^{N_{\rm p}\times 3N_{\rm p}}.
\end{equation}
Using a column index \((k,\alpha)\), corresponding to component \(\alpha\) of particle \(k\), the entries are
\begin{equation}
D_{i,(k,\alpha)} =
\begin{cases}
    d_{ik}^{\alpha},
    & k\in\mathcal N_i,\quad k\neq i,\\[4pt]
    -\displaystyle\sum_{j\in\mathcal N_i}d_{ij}^{\alpha},
    & k=i,\\[8pt]
    0,
    & \text{otherwise}.
\end{cases}
\label{eq:D_matrix_entries}
\end{equation}
Thus
\begin{equation}
    (D\Bnp)_i
    =
    \sum_{k=1}^{N_{\rm p}}
    \sum_{\alpha\in\{x,y,z\}}
    D_{i,(k,\alpha)}B_k^{\alpha}.
\end{equation}
This expression explicitly shows that \(D\) is a linear operator acting on the stacked vector \(\Bnp\).

\subsubsection{Adjoint gradient for the Lagrange multiplier}

The Lagrange multiplier \(\pi_i\) was introduced in Section~\ref{subsec:constrained_minimization} as the multiplier enforcing the constraint \(D\Bnp=0\).
As found in equation~(\ref{eq:G_def}), the correction operator is
\begin{equation}
    \G=M_V^{-1}D^T.
    \label{eq:G_matrix_def}
\end{equation}
Using the matrix entries above, the \(\alpha\)-component of this vector at particle \(i\) is
\begin{align}
(\G\pi)_i^{\alpha}
&=
\frac{1}{V_i}(D^T\pi)_{(i,\alpha)} \nonumber\\
&=
\frac{1}{V_i}
\sum_{k=1}^{N_{\rm p}}
D_{k,(i,\alpha)}\pi_k \nonumber\\
&=
\frac{1}{V_i}
\left[
\sum_{\substack{k=1\\ k\neq i,\ i\in\mathcal N_k}}^{N_{\rm p}}
\pi_k d_{ki}^{\alpha}
-
\pi_i
\sum_{\ell\in\mathcal N_i}
d_{i\ell}^{\alpha}
\right] \nonumber\\
&=
\frac{1}{V_i}
\left[
\sum_{k\in\mathcal N_i}
\pi_k d_{ki}^{\alpha}
-
\pi_i
\sum_{\ell\in\mathcal N_i}
d_{i\ell}^{\alpha}
\right] \nonumber\\
&=
\frac{1}{V_i}
\left[
\sum_{j\in\mathcal N_i}
\pi_j d_{ji}^{\alpha}
-
\pi_i
\sum_{j\in\mathcal N_i}
d_{ij}^{\alpha}
\right].
\label{eq:G_action_components}
\end{align}
In the second equality after the full matrix sum, the first term collects rows \(k\) for which \(i\in\mathcal N_k\), while the second term is the diagonal contribution from row \(i\).
We then use the symmetry of the neighbor graph, \(i\in\mathcal N_k \Leftrightarrow k\in\mathcal N_i\), and finally relabel \(k\) as \(j\).
In vector notation, this gives
\begin{equation}
(\G\pi)_i
=
\frac{1}{V_i}
\left[
\sum_{j\in\mathcal N_i}
\pi_j\dji
-
\pi_i
\sum_{j\in\mathcal N_i}
\dij
\right].
\label{eq:G_action}
\end{equation}
Substituting equations (\ref{eq:dij_def}) and (\ref{eq:dji_def}), and using \(V_i=m_i/\rho_i\), gives the explicit componentwise particle-sum expression
\begin{equation}
(\G\pi)_i^{\alpha}
=
-\sum_{j\in\mathcal N_i}
\left[
\pi_i
\frac{m_j}{m_i\Omega_i}
\left[\nabla_i W(|\ri-\rj|,h_i)\right]^{\alpha}
+
\pi_j
\frac{\rho_i}{\Omega_j\rho_j}
\left[\nabla_i W(|\ri-\rj|,h_j)\right]^{\alpha}
\right].
\label{eq:G_action_expanded}
\end{equation}
Here we used \(\nabla_j W(|\rj-\ri|,h_j)=-\nabla_i W(|\ri-\rj|,h_j)\).
The two terms inside the bracket originate from \(\dij\) and \(\dji\), respectively.
This distinction is essential for an SPH divergence operator.

Equation~(\ref{eq:G_action_expanded}) takes a form consistent with the standard symmetric SPH discretization if we introduce a scalar potential \(\psi_i\) by \(\pi_i=V_i\psi_i\):
\[
(\G\pi)_i^{\alpha}
=
-\rho_i
\sum_{j\in\mathcal N_i}
m_j
\left[
\frac{\psi_i}{\Omega_i\rho_i^2}
\left[\nabla_i W(|\ri-\rj|,h_i)\right]^{\alpha}
+
\frac{\psi_j}{\Omega_j\rho_j^2}
\left[\nabla_i W(|\ri-\rj|,h_j)\right]^{\alpha}
\right].
\]
Using the Lagrange multiplier \(\pi_i\) as the primary scalar variable keeps the adjoint relation and the resulting projection operator transparent.

Equation~(\ref{eq:G_matrix_def}) is equivalent to the volume-metric adjoint relation for arbitrary particle vector fields.
Indeed, for any stacked particle vector field \(X\),
\begin{align*}
    \pi^T D X
    &=
    (D^T\pi)^T X \\
    &=
    (M_V^{-1}D^T\pi)^T M_V X \\
    &=
    (\G\pi)^T M_V X,
\end{align*}
where we used equation~(\ref{eq:G_matrix_def}) and the symmetry of the diagonal matrix \(M_V\).
Written as particle sums, this gives the adjoint relation below.
Conversely, if the adjoint relation holds for arbitrary \(X\), then \((D^T\pi)^T X=(M_V\G\pi)^T X\) for all \(X\).
Therefore \(D^T\pi=M_V\G\pi\), which gives equation~(\ref{eq:G_matrix_def}).
The adjoint relation is
\begin{equation}
\sum_i \pi_i(D\bm{X})_i
=
\sum_i V_i(\G\pi)_i\cdot\bm{X}_i
\label{eq:adjoint_identity}
\end{equation}
for any particle scalar field \(\pi_i\) and any particle vector field \(\bm{X}_i\).
Equation~(\ref{eq:adjoint_identity}) is the volume-metric discrete analogue of integration by parts.
The corresponding vector-calculus identity is
\[
\int_{\Omega}\pi\,\nabla\cdot\bm{X}\,\dd V
=
\int_{\partial\Omega}\pi\,\bm{X}\cdot\bm{n}\,\dd S
+
\int_{\Omega}(-\nabla\pi)\cdot\bm{X}\,\dd V .
\]
When the boundary term vanishes, this becomes
\[
\int_{\Omega}\pi\,\nabla\cdot\bm{X}\,\dd V
=
\int_{\Omega}(-\nabla\pi)\cdot\bm{X}\,\dd V .
\]
Thus \(\G\pi\) is the discrete counterpart of \(-\nabla\pi\) associated with the chosen divergence operator \(D\) and the volume metric.

\subsection{Properties of the Volume-Metric Adjoint Projection}

The adjoint relation is essential to the projection: it makes the correction space generated by \(\G\) orthogonal, with respect to the volume inner product, to the discrete divergence-free space \(\ker(D)\).
In this section, we demonstrate this fact through the resulting orthogonal decomposition, the discrete magnetic-energy estimate, and the conjugate-gradient solvability of the scalar projection equation.

\subsubsection{Orthogonality of the projection}

Let
\begin{equation}
    \ker(D)=\{\bm{X}:D\bm{X}=0\}
\end{equation}
be the discrete divergence-free subspace, and
\begin{equation}
    {\rm range}(\G)=\{\G\pi:\pi\ \text{arbitrary}\}
\end{equation}
be the space of projection corrections.
For any \(\bm{X}\in\ker(D)\), the adjoint identity gives
\begin{equation}
    \sum_i V_i(\G\pi)_i\cdot\bm{X}_i
    =
    \sum_i\pi_i(D\bm{X})_i
    =0.
\end{equation}
Thus
\begin{equation}
    {\rm range}(\G)\perp\ker(D)
\end{equation}
with respect to the volume inner product.

The corrected field satisfies
\begin{equation}
    \Bnp^{n+1}\in\ker(D),
\end{equation}
while the removed component \(\G\pi\) lies in \({\rm range}(\G)\).
Therefore
\begin{equation}
    \Bnp^*=\Bnp^{n+1}+\G\pi
\end{equation}
is an orthogonal decomposition in the volume metric.
Consequently, for consistent boundary and nullspace treatment,
\begin{equation}
    \|\Bnp^*\|_V^2
    =
    \|\Bnp^{n+1}\|_V^2
    +
    \|\G\pi\|_V^2,
\end{equation}
where
\begin{equation}
    \|\bm{X}\|_V^2=\sum_i V_i|\bm{X}_i|^2.
\end{equation}
Thus the method is not merely a divergence-error removal prescription; it is the volume-metric orthogonal projection of \(\Bnp^*\) onto the subspace satisfying \(D\Bnp=0\).
Physically, because the volume metric approximates the magnetic energy of the particle field, \(E_B=(8\pi)^{-1}\|\Bnp\|_V^2=(8\pi)^{-1}\sum_i V_i|\B_i|^2\), this orthogonal decomposition also ensures
\begin{equation}
    E_B^{n+1}
    =
    E_B^*
    -
    \frac{1}{8\pi}\|\G\pi\|_V^2
    \le
    E_B^*,
\end{equation}
with equality only when no correction is applied.
Therefore, a numerically important property of the projection is that it does not increase the discrete magnetic energy; divergence removal by this projection does not inject magnetic energy.

\subsubsection{Conjugate-gradient solvability}

Although the SPH derivative coefficients are directed and generally satisfy \(\dij\neq\dji\), the scalar projection equation, \(D\G\pi=r\), i.e., \(D M_V^{-1}D^T\pi=r\), can be solved by the preconditioned conjugate-gradient (PCG) method because the scalar projection operator is symmetric positive semidefinite.
We introduce this operator as
\begin{equation}
    L\equiv D\G = D M_V^{-1}D^T.
    \label{eq:L_def}
\end{equation}
Because \(M_V^{-1}\) is symmetric, \(L\) is also symmetric:
\begin{equation}
    L^T
    =
    \left(D M_V^{-1}D^T\right)^T
    =
    D\left(M_V^{-1}\right)^T D^T
    =
    D M_V^{-1}D^T
    =
    L.
\end{equation}
The operator \(L\) also satisfies, for any vector \(\bm{X}\) in the multiplier space,
\begin{equation}
    \bm{X}^T L\bm{X}
    =
    (D^T\bm{X})^T M_V^{-1}(D^T\bm{X})
    \ge 0.
\end{equation}
Thus \(L\) is symmetric positive semidefinite in the ordinary scalar-product space of the multiplier.
After removing the null space, the residual equation can be solved by PCG.

\subsection{Numerical Implementation and Pseudocode}
\label{sec:numerical_implementation}

\subsubsection{Implementation flow}

Below, we describe the concrete implementation flow of the projection step.
We distinguish two particle sets.
The active set \(\mathcal A\) contains particles whose magnetic field is corrected by the projection and whose scalar multiplier \(\pi_i\) is an unknown variable.
The fixed set \(\mathcal F\) contains particles whose magnetic field is held fixed during the projection, such as boundary particles; for these particles we set \(\pi_i=0\) and \(\B_i^{n+1}=\B_i^*\).
Fixed particles are not used as rows of the projection equation, but they may appear in neighbor lists of active particles as prescribed magnetic-field values.

For each active particle \(i\in\mathcal A\), the neighbor list \(\mathcal N_i\) is built and the directed edge coefficient \(\dij\) is stored or recomputed.
The same coefficient is then used in both the divergence operator and its adjoint.

The unprojected residual is evaluated as
\begin{equation}
    r_i=(D\Bnp^*)_i
    =
    \sum_{j\in\mathcal N_i}
    \dij\cdot(\B_j^*-\B_i^*).
    \label{eq:implementation_residual}
\end{equation}
for \(i\in\mathcal A\).
The neighbor \(j\) may be active or fixed; if it is fixed, \(\B_j^*\) is simply a prescribed value in this row.

For a trial scalar multiplier \(\pi\), the adjoint gradient is
\begin{align}
    (\G\pi)_i
    &=
    (M_V^{-1}D^T\pi)_i \nonumber\\
    &=
    \frac{1}{V_i}
    \left[
    \sum_{\substack{k\in\mathcal A\\ k\neq i,\ i\in\mathcal N_k}}
    \pi_k \bm{d}_{ki}
    -
    \pi_i
    \sum_{j\in\mathcal N_i}
    \dij
    \right] \nonumber\\
    &=
    \frac{1}{V_i}
    \left[
    \sum_{j\in\mathcal N_i}
    \pi_j\dji
    -
    \pi_i
    \sum_{j\in\mathcal N_i}
    \dij
    \right],
    \label{eq:implementation_G}
\end{align}
for \(i\in\mathcal A\), where the first sum in the second line is over active row particles \(k\) whose neighbor list contains \(i\).
The last equality uses the symmetry of the neighbor graph and the convention \(\pi_j=0\) for \(j\in\mathcal F\); equivalently, the \(\pi_j\dji\) contribution is included only for active neighbor particles.

Then \((L\pi)_i\) is obtained by applying the divergence operator once more:
\begin{equation}
    (L\pi)_i
    =
    \sum_{j\in\mathcal N_i}
    \dij\cdot
    \left[
    \bm{g}_j-\bm{g}_i
    \right].
    \label{eq:implementation_L}
\end{equation}
Here \(\bm{g}_i=(\G\pi)_i\) for \(i\in\mathcal A\), while \(\bm{g}_j=0\) is used when \(j\in\mathcal F\).
Thus applying \(L\) requires one adjoint-gradient evaluation and one divergence evaluation; no explicit global matrix for \(L\) is assembled.

The linear system \(L\pi=r\) is solved by PCG on the active set \(\mathcal A\).
The unprojected particle values \(\B_i^*\) are known on both the active particles \(\mathcal A\) and the fixed particles \(\mathcal F\), but the scalar multiplier and the magnetic correction are solved only on \(\mathcal A\).
We introduce the scalar product of stacked scalar vectors on \(\mathcal A\) as
\begin{equation}
    (a,b)=\sum_{i\in\mathcal A} a_i b_i.
\end{equation}
At iteration \(m\), the PCG residual is
\begin{equation}
    s^{(m)}=r-L\pi^{(m)}.
\end{equation}
The diagonal preconditioner \(P_{\pi}\) is formed from the diagonal of \(L\) as
\begin{equation}
    \bm{q}_i=\sum_{j\in\mathcal N_i}\dij,
    \qquad
    a_i=
    \frac{|\bm{q}_i|^2}{V_i}
    +
    \sum_{j\in\mathcal N_i\cap\mathcal A}
    \frac{|\dij|^2}{V_j},
    \qquad
    P_{\pi}={\rm diag}(a_i).
    \label{eq:pcg_preconditioner}
\end{equation}
Here \(i\in\mathcal A\).
The preconditioned residual is then
\begin{equation}
    z_i^{(m)}
    =
    (P_{\pi}^{-1}s^{(m)})_i
    =
    \frac{s_i^{(m)}}{a_i+\epsilon_a},
    \label{eq:pcg_preconditioned_residual}
\end{equation}
where \(\epsilon_a\) is a small regularization used only if needed.
The second sum in equation~(\ref{eq:pcg_preconditioner}) includes only active neighbors.
For the PCG recurrence, we define
\begin{equation}
    \gamma_m=(s^{(m)},z^{(m)}).
\end{equation}
The search direction is initialized and updated as
\begin{equation}
    p^{(0)}=z^{(0)},
    \qquad
    p^{(m)}=z^{(m)}+\beta_{m-1}p^{(m-1)}
    \quad (m\ge 1),
    \qquad
    \beta_{m-1}=\frac{\gamma_m}{\gamma_{m-1}}.
    \label{eq:pcg_search_direction}
\end{equation}
Given the search direction \(p^{(m)}\), the unfiltered operator action is evaluated as
\begin{equation}
    \bm{g}^{(m)}=\G p^{(m)},
    \qquad
    \tilde w^{(m)}
    =
    D\bm{g}^{(m)}
    =
    L p^{(m)}.
    \label{eq:cg_operator_action}
\end{equation}
When mean subtraction is required, for example to remove the constant scalar null mode in periodic calculations, the unfiltered scalar action \(\tilde w^{(m)}\) in equation (\ref{eq:cg_operator_action}) is replaced before the PCG update by
\begin{equation*}
    w_i^{(m)}
    =
    (P_{\rm c}\tilde w^{(m)})_i
    =
    \tilde w_i^{(m)}-\bar{\tilde w}^{(m)}_V,
    \qquad i\in\mathcal A.
    \tag{\ref{eq:cg_operator_action}$'$}
\end{equation*}
Here \(P_{\rm c}\) denotes the scalar mean-subtraction operator and
\begin{equation}
    \bar{\tilde w}^{(m)}_V
    =
    \frac{\sum_{k\in\mathcal A}V_k \tilde w_k^{(m)}}
         {\sum_{k\in\mathcal A}V_k}.
    \label{eq:Pc_def}
\end{equation}
For cases where mean subtraction is not required, \(w^{(m)}=\tilde w^{(m)}\), equivalently \(P_{\rm c}\) is the identity.
In the code this optional step corresponds to subtracting the volume-weighted mean from the scalar operator-action vector in the periodic case.
In the tests presented in this paper, mean subtraction is used only for the Dedner-type divergence test with periodic boundaries.

\begin{table}[t]
\caption{Preconditioned conjugate-gradient projection algorithm.}
\label{tab:cg_projection_algorithm}
\begin{algorithmic}[1]
\Require \(\B_i^*\) on \(\mathcal A\cup\mathcal F\), neighbor lists \(\mathcal N_i\), and directed coefficients \(\dij\)
\State Compute \(r\gets D\Bnp^*\) on \(\mathcal A\)
\State Compute the diagonal preconditioner \(P_{\pi}={\rm diag}(a_i)\)
\State Set \(\pi^{(0)}\gets0\) and \(s^{(0)}\gets r\)
\State Set \(z^{(0)}\gets P_{\pi}^{-1}s^{(0)}\) and \(p^{(0)}\gets z^{(0)}\)
\State Set \(\gamma\gets(s^{(0)},z^{(0)})\)
\For{\(m=0,1,\ldots\)}
    \State \(\bm{g}^{(m)}\gets\G p^{(m)}\)
    \State \(\tilde w^{(m)}\gets D\bm{g}^{(m)}=L p^{(m)}\)
    \State \(w^{(m)}\gets\tilde w^{(m)}\), or \(P_{\rm c}\tilde w^{(m)}\) when mean subtraction is used
    \State \(\displaystyle \alpha_m\gets\frac{\gamma}{(p^{(m)},w^{(m)})}\)
    \State \(\pi^{(m+1)}\gets\pi^{(m)}+\alpha_m p^{(m)}\)
    \State \(s^{(m+1)}\gets s^{(m)}-\alpha_m w^{(m)}\)
    \If{the stopping criterion is satisfied}
        \State \textbf{break}
    \EndIf
    \State \(z^{(m+1)}\gets P_{\pi}^{-1}s^{(m+1)}\)
    \State \(\gamma_{\rm new}\gets(s^{(m+1)},z^{(m+1)})\)
    \State \(\displaystyle \beta_m\gets\frac{\gamma_{\rm new}}{\gamma}\)
    \State \(p^{(m+1)}\gets z^{(m+1)}+\beta_m p^{(m)}\)
    \State \(\gamma\gets\gamma_{\rm new}\)
\EndFor
\State Set \(\B_i^{n+1}\gets \B_i^*-(\G\pi)_i\) for \(i\in\mathcal A\)
\State Set \(\B_i^{n+1}\gets \B_i^*\) for \(i\in\mathcal F\)
\end{algorithmic}
\end{table}

\subsubsection{Convergence criterion used in this paper}

At PCG iteration \(m\), the provisional corrected magnetic field is
\begin{equation}
    \Bnp^{(m)}=\Bnp^*-\G\pi^{(m)}.
\end{equation}
Since \(r=D\Bnp^*\) and \(L=D\G\), the residual of the projection equation is
\begin{equation}
    r-L\pi^{(m)}
    =
    D\Bnp^{(m)}.
\end{equation}
In a periodic domain, the constant mode of the residual is removed before solving the projection equation.
We therefore write the convergence measures in terms of the mean-free discrete divergence of the provisional field.
For a particle vector \(q\), we define this mean-subtraction operator by
\begin{equation}
    (P_{\rm c}q)_i
    =
    q_i
    -
    \frac{\sum_{j=1}^{N_{\rm p}} V_j q_j}{\sum_{j=1}^{N_{\rm p}} V_j}.
    \label{eq:Pc_definition}
\end{equation}
Thus the monitored residual is \(P_{\rm c}D\Bnp^{(m)}\), not the raw \(D\Bnp^{(m)}\).
A standard relative solver tolerance may be written as
\begin{equation}
    \|r_m\|\equiv\|P_{\rm c}D\Bnp^{(m)}\|_V,
    \qquad
    \frac{\|r_m\|}{\|P_{\rm c}D\Bnp^*\|_V+\epsilon}<\epsilon_{\rm rel},
    \qquad
    \|D\Bnp\|_V^2=\sum_i V_i\left[(D\Bnp)_i\right]^2.
\end{equation}
This relative residual criterion is used for the periodic Dedner-type divergence tests presented below.
For non-periodic applications, one may simply set \(P_{\rm c}=I\).

On the other hand, for the collapse test in this paper, we use a more practical stopping criterion based on the divergence error accumulated during the interval between two projection steps.
This criterion consists of two conditions.
The first is a top-tail condition, which is satisfied either by reducing the divergence change accumulated since the previous projection or by making the current top-tail divergence sufficiently small.
The second is an all-particle absolute condition.

We define the dimensionless divergence measure
\begin{equation}
    \chi_i^{(m)}
    =
    \left|
    \frac{(D\Bnp^{(m)})_i}{|\B_i^{(m)}|/h_i}
    \right|.
\end{equation}
We store \(\chi_i^{\rm prev}\), the value just after the previous projection.
The change in the divergence measure during the current projection interval is then measured by
\begin{equation}
    \Delta\chi_i^{(m)}
    =
    \chi_i^{(m)}-\chi_i^{\rm prev}.
\end{equation}
Let \({\rm RMS}_{f_{\rm top}}(\Delta\chi)\) denote the RMS of the largest fraction \(f_{\rm top}\) of \(|\Delta\chi_i|\) over particles in \(\mathcal A\), and define \({\rm RMS}_{f_{\rm top}}(\chi)\) analogously.
The notation \({\rm RMS}(x)\), without a subscript, denotes the ordinary RMS over \(\mathcal A\).
The first condition for the top-tail particles is
\begin{equation}
    {\rm RMS}_{f_{\rm top}}(\Delta\chi^{(m)})
    \le
    f_{\rm red}
    {\rm RMS}_{f_{\rm top}}(\Delta\chi^{(0)})
    \quad
    \text{or}
    \quad
    {\rm RMS}_{f_{\rm top}}(\chi^{(m)})<\epsilon_{\rm abs},
    \label{eq:collapse_increment_stop}
\end{equation}
where \(f_{\rm top}\) specifies the fraction of particles used in the top-tail RMS, and \(f_{\rm red}\) specifies the required reduction factor.
In the calculations presented here we use \(f_{\rm top}=0.01\) and \(f_{\rm red}=0.1/N_{\rm interval}\), where \(N_{\rm interval}\) is the number of ordinary MHD updates between two projection steps.
Thus the reduction condition in equation~(\ref{eq:collapse_increment_stop}) continues the iterations until the divergence accumulated by the ordinary MHD update, measured over the largest 1\% of particles, is reduced to one tenth of the average amount accumulated in one ordinary MHD update during the preceding projection interval.
The top-tail absolute condition prevents unnecessary iteration when the current top-tail divergence is already below the absolute tolerance.
The second condition is
\begin{equation}
    {\rm RMS}(\chi^{(m)})<\epsilon_{\rm abs}.
    \label{eq:collapse_absolute_stop}
\end{equation}
This condition requires the ordinary RMS over all active particles to be below the absolute tolerance.
The projection is stopped only when both equations~(\ref{eq:collapse_increment_stop}) and (\ref{eq:collapse_absolute_stop}) are satisfied.
The strategy for the collapse test is therefore to reduce the divergence that has accumulated since the last projection.
It does not aim to over-solve the projection far below the typical level of \(D\B\) generated by the underlying MHD scheme, for example down to machine precision, although such a tighter solve is achievable with sufficiently many iterations.

\section{Test calculations}

\subsection{Dedner-type divergence test}

We first test our projection method with a two-dimensional periodic divergence test based on \citet{2002JCoPh.175..645D}.
The calculation is performed in a unit square with \(64^2\) equal-mass particles.
Instead of evolving the Dedner cleaning equations, we use their compact magnetic perturbation as the magnetic field \(\B^*\) and apply the projection once.
The imposed field is centered at \((x,y)=(1/2,1/2)\) and is
\begin{equation}
    B_x^*
    =
    q^8-2q^4+1,
    \qquad
    B_y^*=B_z^*=0
    \quad
    (q\le 1),
\end{equation}
where \(q=r/r_0\), \(r\) is the periodic distance from the center, and \(r_0=0.20\).
For \(q>1\), we set \(\B^*=0\).
This field is compactly supported and smooth at \(q=1\), but it has a finite discrete \(D\B\) because \(B_x^*\) varies in space.

We present results for two particle arrangements.
The first is a Cartesian lattice randomly displaced by 10 per cent of the grid spacing in each coordinate.
We use this perturbed lattice, rather than an exact Cartesian lattice, because in the exact periodic lattice the residual did not decrease cleanly, apparently owing to symmetry and periodic-boundary effects.
The second is a random distribution sampled uniformly over the simulation domain.
In both cases, \(\rho_i\) and \(h_i\) are iterated self-consistently from the SPH density relation.
Thus the imposed Dedner-type magnetic perturbation is tested at the same nominal \(64^2\) resolution with different levels of particle disorder.

\begin{figure}
\centering
\includegraphics[width=0.96\textwidth]{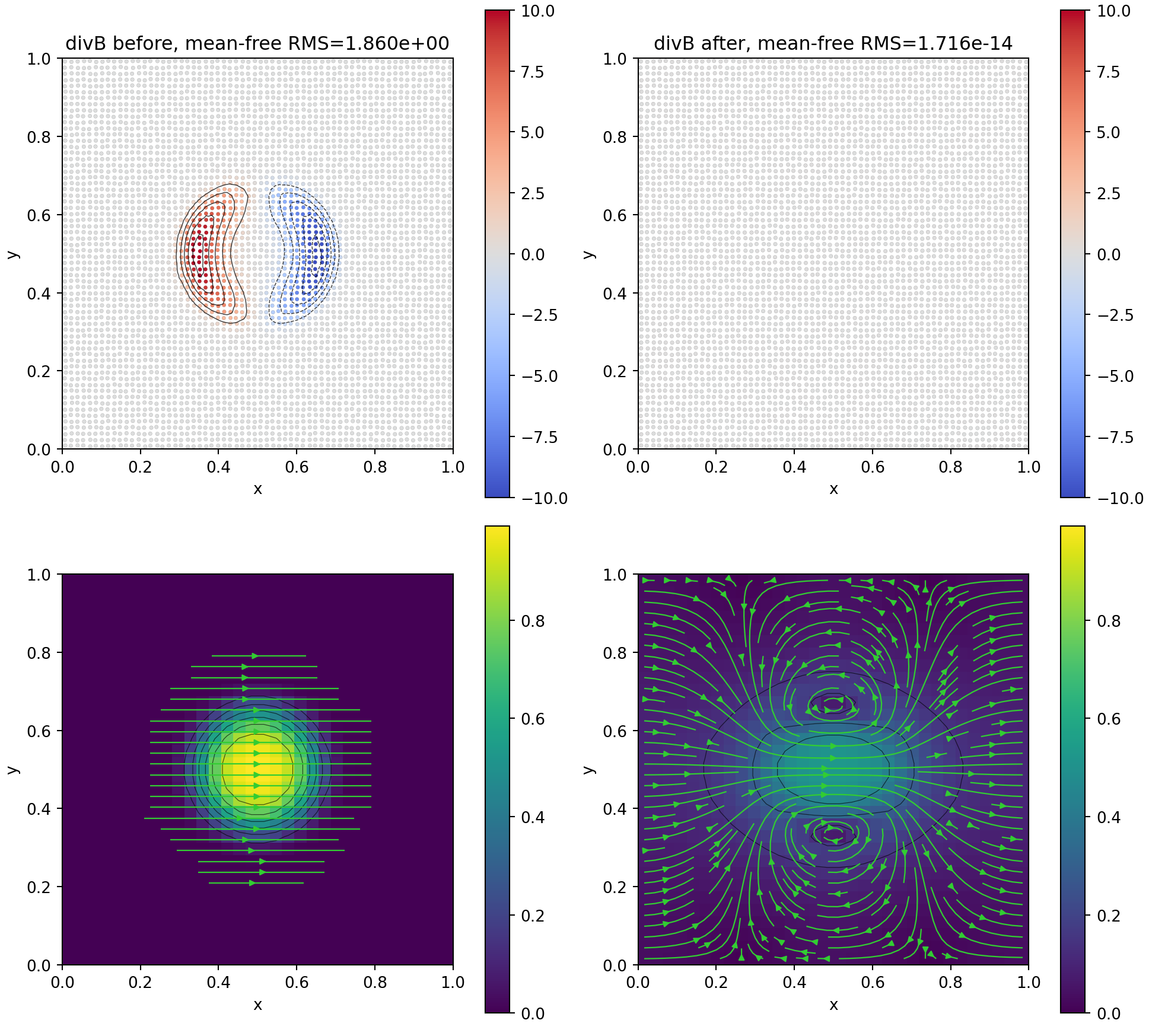}
\caption{Dedner-type projection test on a Cartesian particle lattice with a random displacement of 10 per cent of the grid spacing in each coordinate.
The upper panels show the mean-free discrete divergence at the particle positions before and after projection; black contours show the contours of \(|D\B|=2,4,6,8,10\).
The lower panels show the magnetic-field strength and magnetic field lines before and after the correction.}
\label{fig:dedner_grid_disp10}
\end{figure}

\begin{figure}
\centering
\includegraphics[width=0.96\textwidth]{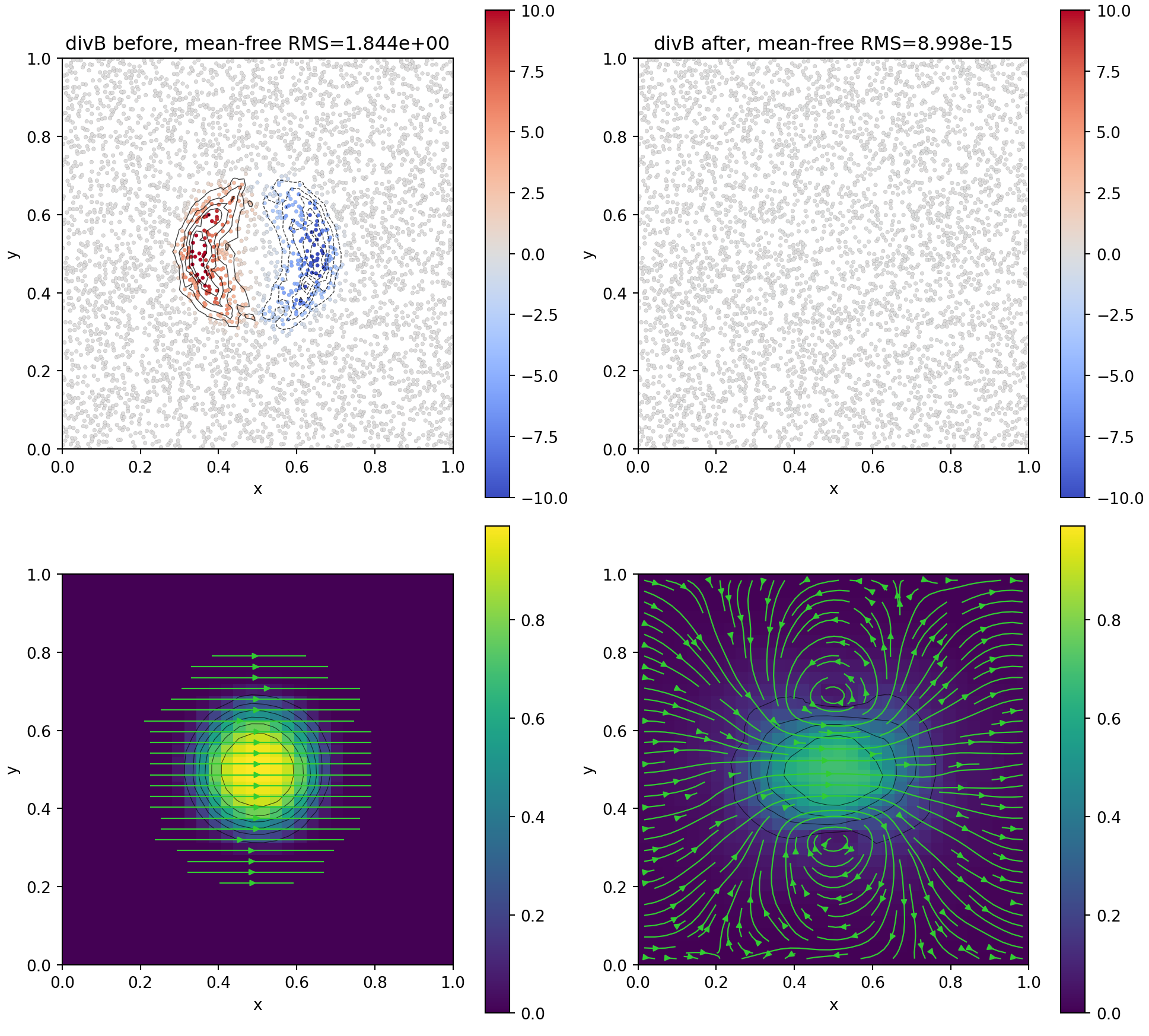}
\caption{Same Dedner-type divergence test as in Figure \ref{fig:dedner_grid_disp10}, but on a random particle distribution.
The calculation uses \(64^2\) uniformly sampled particles.}
\label{fig:dedner_random}
\end{figure}

Figure \ref{fig:dedner_grid_disp10} shows the result for the 10 per cent displaced lattice.
In the upper-left panel, the imposed field produces a localized discrete divergence error with a maximum amplitude of about \(|D\B|\simeq 11\).
After the projection, the upper-right panel shows that this mean-free divergence is barely visible.
The lower panels show that the magnetic-field strength is reduced near the center and that the corrected field lines have a divergence-free pattern.
This final magnetic configuration should not be over-interpreted physically, because the initial field is an artificial divergence-error pattern rather than a physical magnetic field.
It is nevertheless useful as a visual check of the projection.
As seen below, the corrected field geometry differs slightly for the random particle distribution.

Figure \ref{fig:dedner_random} shows the same test on the random particle distribution.
Because of the initial particle randomness, the contours of \(D\B\) in the upper-left panel are more irregular than in Figure \ref{fig:dedner_grid_disp10}.
However, this is not a problem for the test.
Note that, in practical SPH simulations, particle distributions are often less irregular because pressure forces and other dynamics improve the local ordering.
The purpose of this distribution is to verify that the projection works for a particle set with little apparent symmetry, providing a stringent stress test for the discrete adjoint construction.
As shown in the upper-right panel, the mean-free divergence is again reduced to a sufficiently small value.
The corrected magnetic-field strength and field-line pattern are also broadly consistent with Figure \ref{fig:dedner_grid_disp10}.
These tests indicate that the projection performs well as long as the particle arrangement does not impose an excessive exact symmetry, which does not arise in practical simulations.

\begin{figure}
\centering
\includegraphics[width=0.68\textwidth]{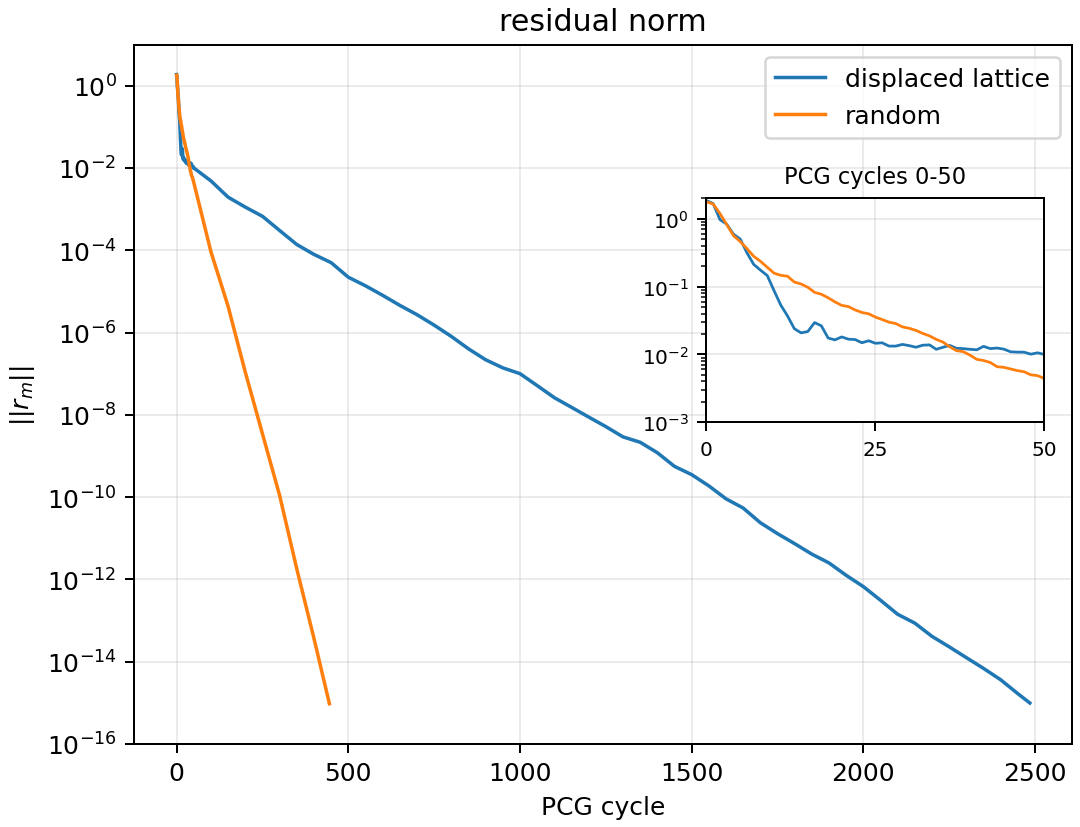}
\caption{Residual convergence of the projection calculations for the Dedner-type divergence tests.
The figure shows the volume-weighted mean-free residual norm \(\|r_m\|=\|P_{\rm c}D\B^{(m)}\|_V\) as a function of the PCG cycle.
The blue and orange curves show the 10 per cent displaced lattice and the random particle distribution, respectively.
The upper-right zoomed panel shows the first 50 PCG cycles.}
\label{fig:dedner_residual_convergence}
\end{figure}

Figure \ref{fig:dedner_residual_convergence} shows the corresponding PCG convergence for the two particle distributions.
Here \(r_m\) denotes the scalar residual of the projection equation after \(m\) PCG iterations, with the periodic mean mode subtracted.
The 10 per cent displaced lattice reaches \(\|r_m\|\sim 10^{-15}\) after about \(2500\) PCG cycles, while the random distribution reaches a comparable level after about \(500\) PCG cycles.
This demonstrates that the present discretely adjoint formulation can reduce the discrete \(D\B\) residual to the level of floating-point roundoff.
This result is noteworthy compared with the projection formulation of \citet{2005MNRAS.364..384P}: they solved the Poisson equation for the correction field by a gravity-like particle summation and found that correcting errors at wavelengths comparable to, or shorter than, the smoothing length was difficult.
Our result indicates that such \(h\)-scale and particle-scale components of \(D\B\) can also be corrected down to the floating-point roundoff level.

It is interesting that the random distribution converges faster than the 10 per cent displaced lattice.
This may seem counterintuitive because the random distribution is expected to produce more irregular discrete matrices \(D\) and \(\G\).
Note, however, that the PCG convergence is controlled by the spectrum of \(L\), the preconditioner, and how the initial residual projects onto the preconditioned eigenmodes, rather than by visual regularity alone.
A plausible explanation is that the nearly periodic lattice still retains Fourier-like coherent modes associated with the periodic symmetry, so that part of the residual lies along slowly converging small-eigenvalue directions.
The random distribution breaks this coherence and makes the imposed localized residual less aligned with the slow modes of \(L\).
This is also consistent with the behavior of the exact periodic lattice, for which the residual did not decrease cleanly.

Note, however, that the upper-right zoomed panel of Figure \ref{fig:dedner_residual_convergence} shows that, during the first 50 PCG cycles, the residual decreases more rapidly for the displaced-lattice case than for the random-particle case.
This suggests that the small-scale residual components are removed more efficiently in a lattice-like particle arrangement.

One might worry that \( \sim 10^3\) PCG cycles are too expensive for practical simulations.
However, the calculations in Figure \ref{fig:dedner_residual_convergence} were intentionally driven to roundoff-level residuals as diagnostic tests; such roundoff-level residuals are not required for practical simulations.
In actual use, the projection only needs to reduce \(D\B\) to the level of the divergence error generated by the MHD solver during the interval between projections; we show below that substantially fewer iterations are sufficient for this purpose.
The projection also need not be applied every timestep.
As long as \(D\B\) remains small enough not to affect the solution, applying the projection every several timesteps is acceptable.
This flexibility is an advantage of solving a time-independent constraint equation, which can be applied to the accumulated divergence error when needed.

\subsection{Collapse calculation: divergence cleaning and projection}

We next apply the projection method to a three-dimensional non-ideal MHD gravitational collapse calculation based on the initial condition of \citet{2023PASJ...75..835T}.
The initial cloud is a rotating, magnetized dusty Bonnor--Ebert sphere, and the mass within the sphere is \(1\,M_\odot\).
The initial cloud has \(\alpha=0.4\), \(\beta=0.03\), and normalized mass-to-flux ratio \(\mu=5\).
The \(1\,M_\odot\) sphere is represented by \(N_{\rm SPH}=3\times10^6\) SPH particles.
In these test calculations, we do not introduce sink particles.
We omit the remaining details of the thermodynamics, resistivity model, and dust treatment because they are not essential for the present divergence-control test.
We compare a divergence-cleaning calculation and projection calculations that use the same initial condition and MHD update.
The divergence-cleaning calculation uses the hyperbolic cleaning prescription described below, while the projection calculations apply the elliptic projection at prescribed intervals.

For the projection calculation, we perform three projection-interval cases with \(N_{\rm interval}=1,10,\) and \(100\).
In all cases, we fix the monitored top fraction to \(f_{\rm top}=0.01\), the increment-reduction factor to \(f_{\rm red}=0.1/N_{\rm interval}\), and the absolute tolerance for \(\chi \equiv h|\nabla\cdot\B|/|\B|\) to \(\epsilon_{\rm abs}=10^{-5}\).
This choice is designed to reduce the mean divergence error generated per MHD update to roughly one tenth of its accumulated value, rather than to solve the elliptic equation to roundoff accuracy.
Specifically, the projection requires the top \(1\%\) RMS of the newly accumulated \(\chi\) increment to be reduced by \(f_{\rm red}\), or the current top \(1\%\) RMS of \(\chi\) to be below \(10^{-5}\), together with the ordinary RMS of \(\chi\) over all active particles being below \(10^{-5}\).

For the divergence-cleaning calculation, we keep the hyperbolic cleaning speed fixed at \(c_h=2.5\times10^2\,{\rm m\,s^{-1}}\).
The damping parameter is \(\sigma_{\rm hdc}=1.0\), so the local damping time is \(\tau_{\rm hdc}=\Delta x_{\rm hdc}/(\sigma_{\rm hdc} c_h)=\Delta x_{\rm hdc}/c_h\).
Here \(\Delta x_{\rm hdc}\) is set to \(h_{\rm min}\), the smallest smoothing length in the active particle set, and therefore represents the smallest resolved length scale of the calculation at that time.
In this setup \(c_h\) is fixed, while \(h_{\rm min}\) is updated with the flow, and therefore \(\tau_{\rm hdc}\) varies in time but is spatially constant at each time.

\begin{figure}
\centering
\includegraphics[width=0.86\textwidth]{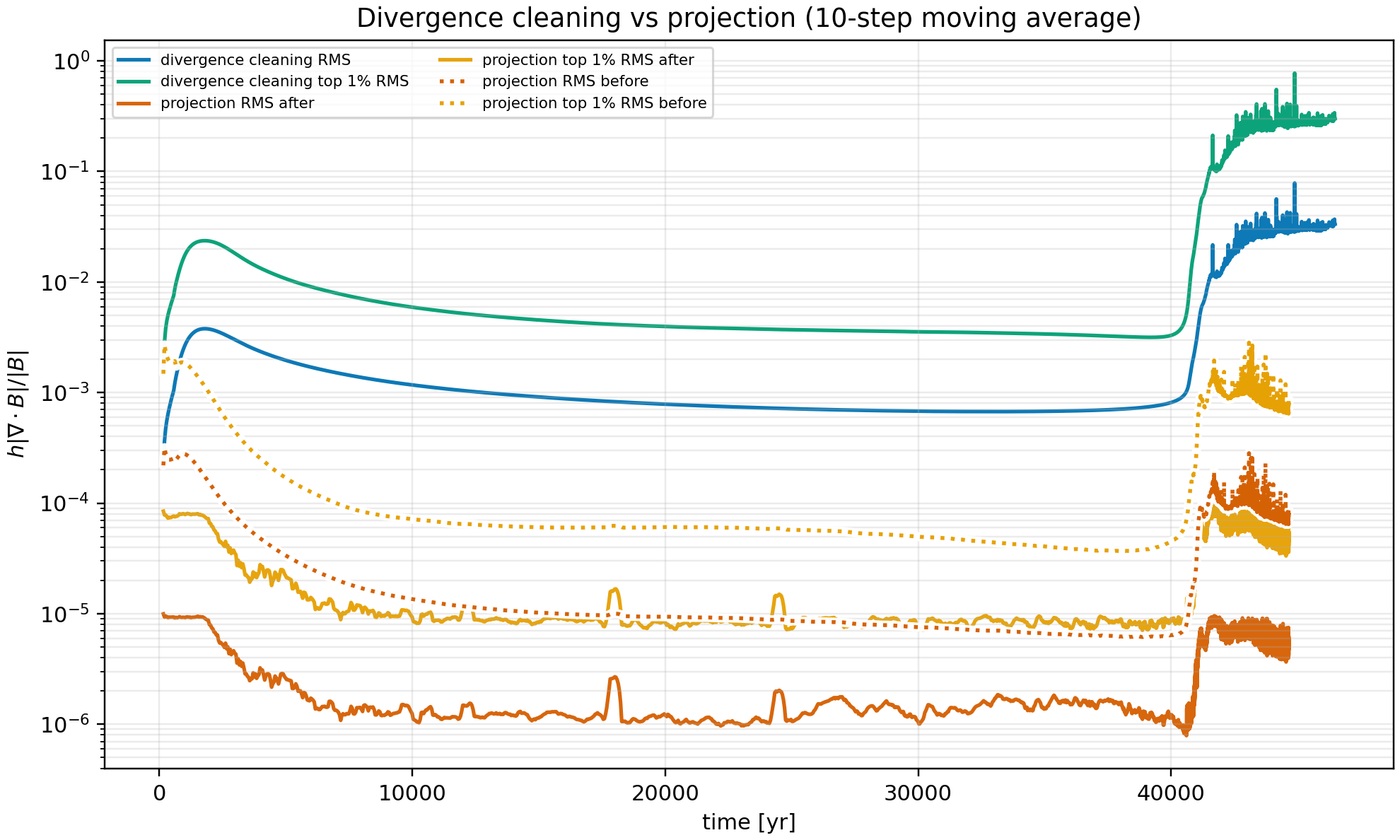}
\caption{Time evolution of the dimensionless divergence error in the magnetized collapse calculation.
The figure compares the run with divergence cleaning and the run with the projection applied every MHD update (\(N_{\rm interval}=1\)).
The plotted quantities are 10-step moving averages of the RMS and top \(1\%\) RMS of \(h|\nabla\cdot\B|/|\B|\).
For the projection run, solid curves show the values immediately after projection, while dotted curves show the values immediately before projection.}
\label{fig:kyoto1_vs_kyoto2_divB}
\end{figure}

Figure \ref{fig:kyoto1_vs_kyoto2_divB} shows the normalized divergence error \(\chi=h|\nabla\cdot\B|/|\B|\) in the collapse calculations.
For the projection run with \(N_{\rm interval}=1\), the solid curves show the values immediately after projection, while the dotted curves show the values immediately before projection, i.e., after one MHD update.
The orange and yellow curves denote the RMS over all active particles and the top \(1\%\) RMS, respectively.
During the isothermal collapse phase, \(t<4\times10^4\,{\rm yr}\), the projection efficiently reduces the divergence error and keeps the global RMS below \(10^{-5}\) and the top \(1\%\) RMS below \(10^{-4}\)
even after one MHD update.
By contrast, in the divergence-cleaning run, the normalized divergence error is larger by more than an order of magnitude during the same phase.

The divergence error subsequently grows when the first adiabatic core forms, at \(t\simeq4.1\times10^4\,{\rm yr}\).
A shock forms at the surface of the first core, and the divergence error appears to be generated mainly around this shocked region.
In the projection run, both the global RMS and the top \(1\%\) RMS increase by more than an order of magnitude, but they are still maintained at \(\sim10^{-4}\) and \(\sim10^{-3}\), respectively.
In the divergence-cleaning run, these quantities also increase and reach \(\sim10^{-2}\) and \(\sim10^{-1}\), respectively, about two orders of magnitude higher than in the projection run.
This late-time increase of \(\nabla\cdot\B\), as well as the magnitude reached in the divergence-cleaning run, is broadly consistent with the divergence-cleaning calculations of \citet{2012JCoPh.231.7214T}.
Although some improvement may be possible by tuning the cleaning parameters, we do not pursue this issue further because it is outside the scope of the present paper.
This comparison shows that the projection removes numerical divergence error much more efficiently than divergence cleaning in the collapse calculation.

To identify a suitable projection interval, we performed additional tests with \(N_{\rm interval}=10\) and 100.
For these interval tests, the increment-reduction threshold is set to \(0.1/N_{\rm interval}\), so the intended reduction of the divergence error generated per MHD update is \(0.1\) in all cases.

\begin{figure}
\centering
\includegraphics[width=0.86\textwidth]{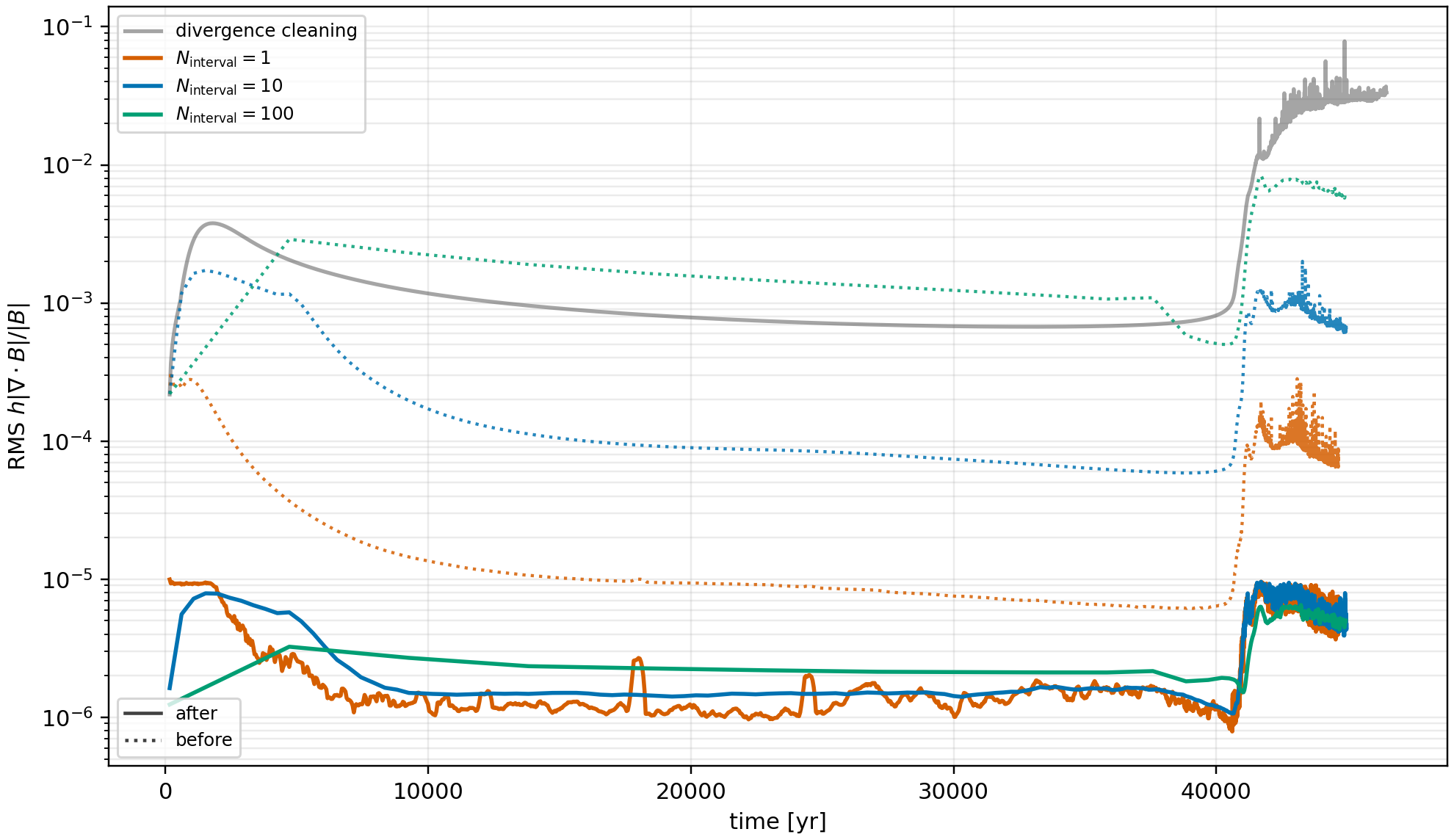}
\caption{Dependence of the normalized divergence RMS error on the projection interval.
The colored curves show projection runs with \(N_{\rm interval}=1\), 10, and 100, while the gray curve shows the divergence-cleaning run.
All plotted curves are 10-step moving averages.
For the projection runs, dotted and solid curves show the values immediately before and after projection, respectively.}
\label{fig:projection_interval_compare_divB_rms}
\end{figure}

Figure \ref{fig:projection_interval_compare_divB_rms} shows the results for \(N_{\rm interval}=1\), 10, and 100.
As shown by the solid curves, the projection reduces the RMS normalized divergence error to \(\lesssim10^{-5}\) for all values of \(N_{\rm interval}\).
However, the pre-projection divergence error, measured immediately before the next projection, i.e., after \(N_{\rm interval}\) MHD updates, increases approximately linearly with \(N_{\rm interval}\).
This behavior indicates that the divergence error generated at each MHD update accumulates almost linearly between projection events.
In our setup, \(N_{\rm interval}=100\) run shows RMS divergence error to become comparable to that in the divergence-cleaning calculation.
A comparison using the top \(1\%\) RMS gives values that are typically about an order of magnitude larger overall, but shows the same trend with \(N_{\rm interval}\).

\begin{figure}
\centering
\includegraphics[width=0.78\textwidth]{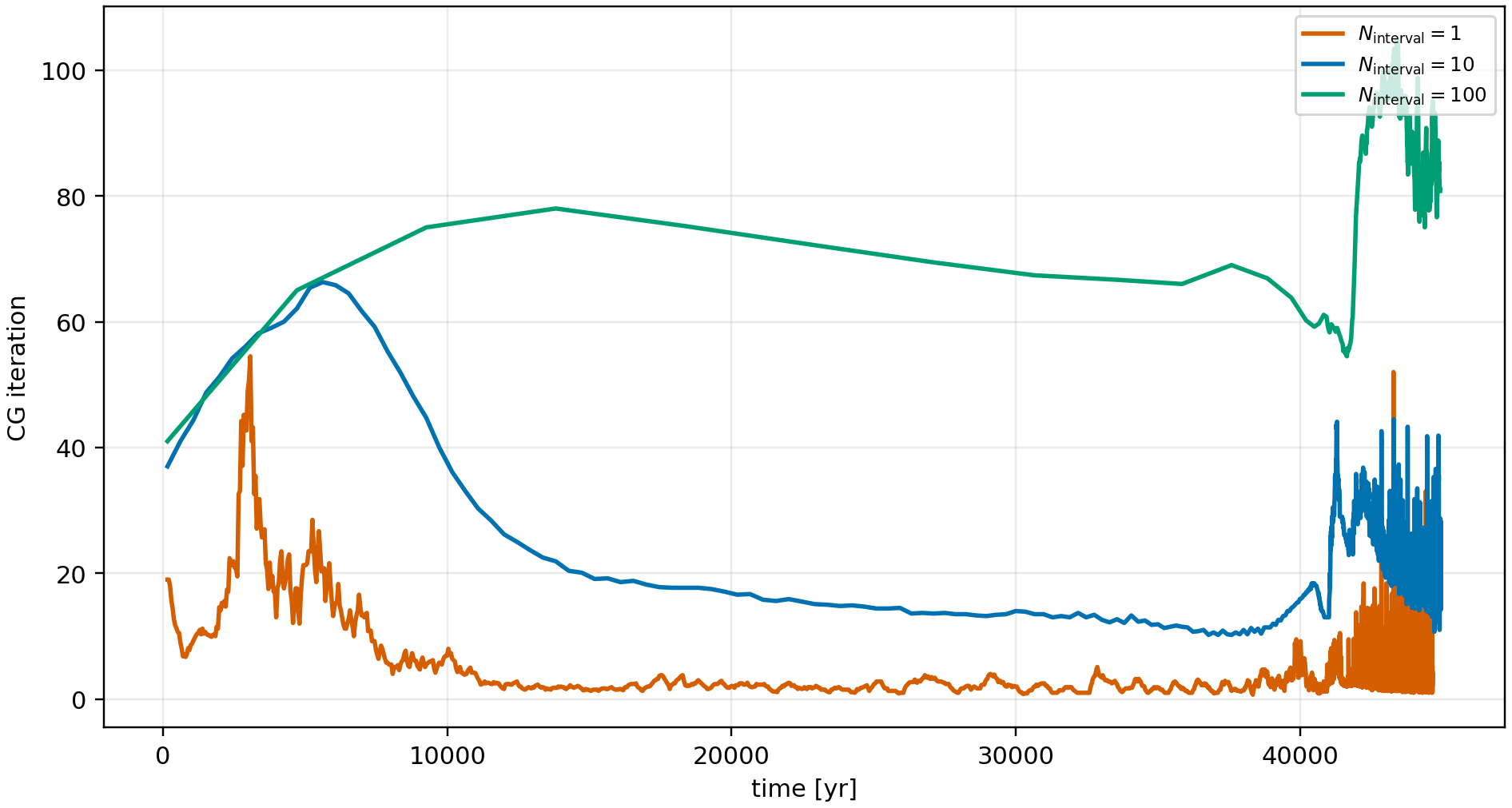}
\caption{Number of PCG iterations required by the projection runs with different projection intervals.
All plotted numbers are 10-step moving averages.}
\label{fig:projection_interval_compare_iteration}
\end{figure}

Figure \ref{fig:projection_interval_compare_iteration} shows the number of PCG iterations required in the projection runs with \(N_{\rm interval}=1\), 10, and 100.
The number of PCG iterations clearly increases as \(N_{\rm interval}\) becomes larger.
This trend is expected because a larger number of MHD updates between projections allows the solution to move farther away from the divergence-free manifold.
Therefore, increasing \(N_{\rm interval}\) does not reduce the numerical cost linearly.

\begin{figure}
\centering
\includegraphics[width=0.78\textwidth]{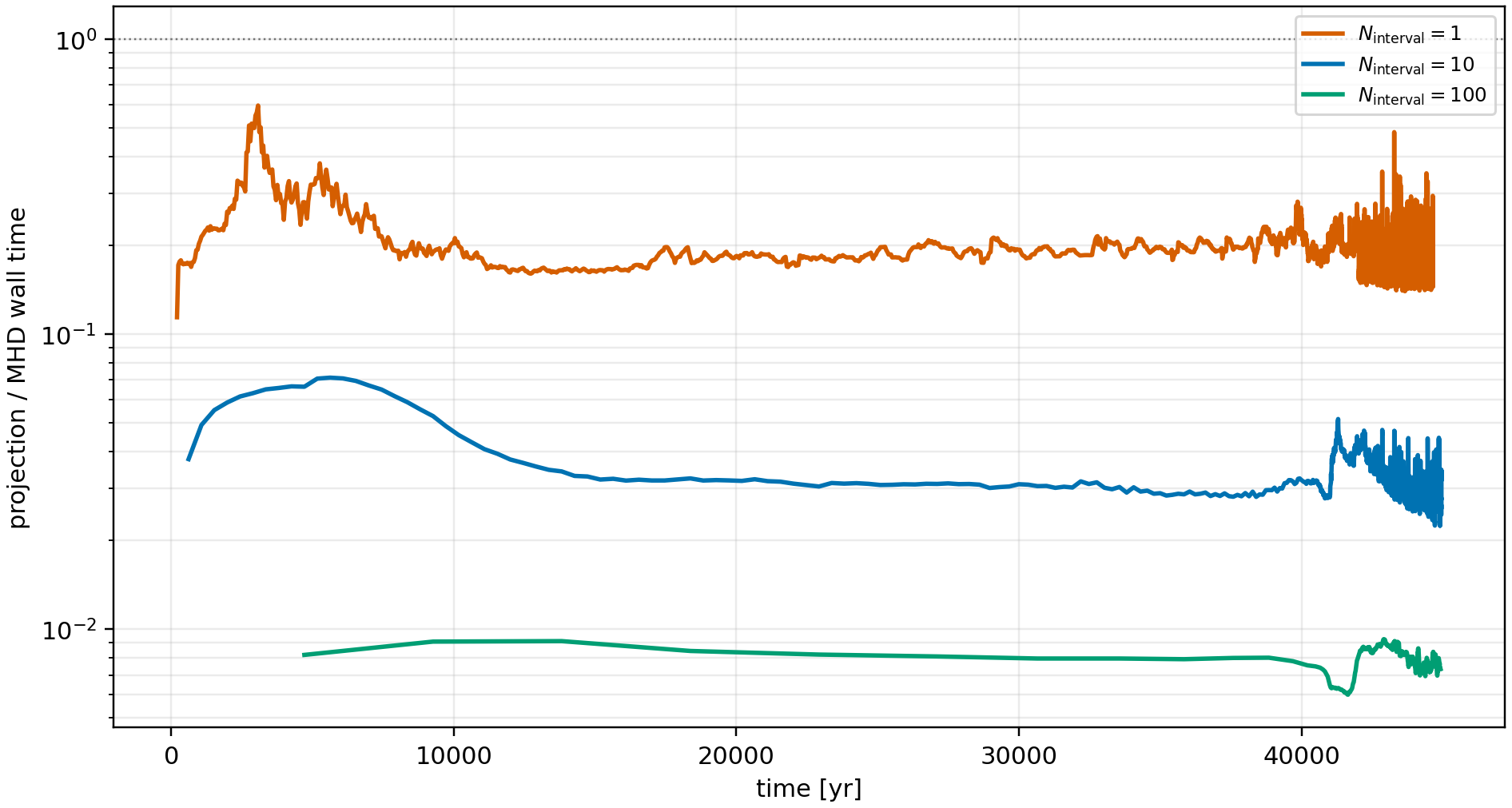}
\caption{Ratio of the projection wall time to the MHD wall time for different projection intervals.
All plotted ratios are 10-step moving averages.}
\label{fig:projection_interval_compare_timing_ratio}
\end{figure}

Figure \ref{fig:projection_interval_compare_timing_ratio} shows, for the runs with \(N_{\rm interval}=1\), 10, and 100, the ratio between the wall time of one projection step and the total wall time spent on the MHD updates over the interval between projections in our code implementation.
This ratio depends on implementation details and optimization, so it should be regarded as an order-of-magnitude guide rather than a machine-independent performance estimate.
The calculation shown here was performed with 128 MPI processes.
In our implementation, the projection cost is about \(20\)--\(60\%\) of the MHD update cost for \(N_{\rm interval}=1\), about \(3\)--\(7\%\) for \(N_{\rm interval}=10\), and below \(1\%\) for \(N_{\rm interval}=100\).
On the other hand, Figure \ref{fig:projection_interval_compare_divB_rms} indicates that \(N_{\rm interval}\lesssim10\) is desirable if the normalized divergence error \(h|\nabla\cdot\B|/|\B|\) is to be kept about the \(0.1\%\) level.
Combining the divergence-error control and the computational cost, \(N_{\rm interval}\lesssim10\) appears to provide a good balance for this setup.

\begin{figure}
\centering
\includegraphics[width=0.78\textwidth]{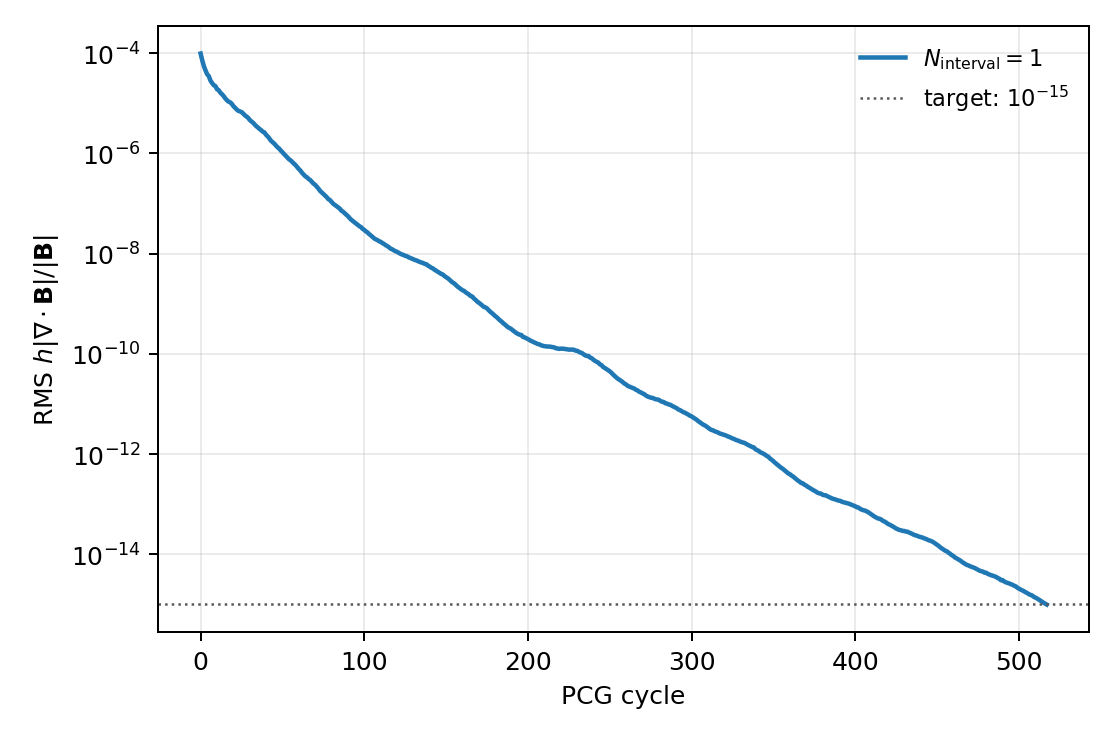}
\caption{Divergence-error convergence test for the magnetized collapse simulation.
The calculation was restarted from a snapshot of the \(N_{\rm interval}=1\) run at \(t\simeq4.2\times10^4\,{\rm yr}\), and only the projection module was iterated until the RMS normalized divergence error reached the roundoff-level target.}
\label{fig:check_divB_to_precision_collapse}
\end{figure}

Figure \ref{fig:check_divB_to_precision_collapse} shows a stress test of the projection solve using a snapshot from the \(N_{\rm interval}=1\) collapse run at \(t\simeq4.2\times10^4\,{\rm yr}\).
Starting from this snapshot, we restarted the calculation and applied only the projection module, continuing the PCG iterations until the RMS normalized divergence error reached the roundoff-level target.
The purpose of this test is to examine whether the projection scheme can still reduce the divergence error to roundoff level in a realistic three-dimensional collapse simulation.
For the target value \(10^{-15}\), the RMS divergence error reaches the target after about 500 PCG cycles.
This result demonstrates that, even in a complex three-dimensional collapse calculation, the present projection scheme can in principle remove numerical divergence error down to the floating-point roundoff level.

\begin{figure*}
\centering
{\setlength{\tabcolsep}{1pt}
\def\xzpanel#1{\includegraphics[width=0.35\textwidth,trim=24 6 40 22,clip]{#1}}
\def\runlabel#1{\raisebox{0.167\textwidth}[0pt][0pt]{\rotatebox[origin=c]{90}{\scriptsize #1}}}
\begin{tabular}{@{}r@{\hspace{0.4em}}c@{\hspace{0.5em}}c@{}}
 & \scriptsize density & \scriptsize plasma \(\beta\) \\[-0.3ex]
\runlabel{cleaning} &
\xzpanel{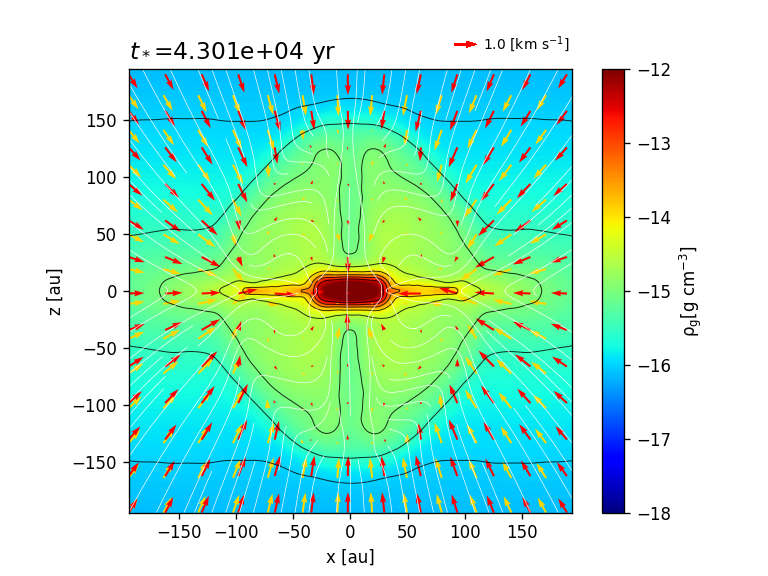} &
\xzpanel{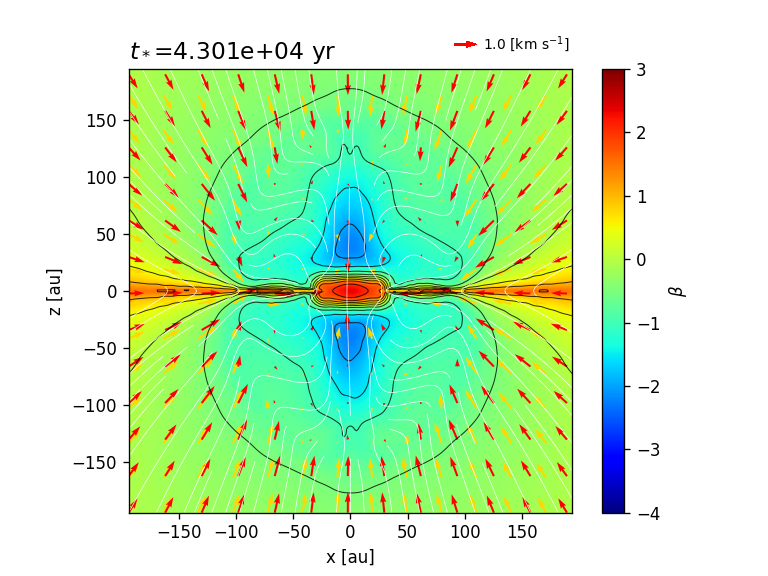} \\[-0.4ex]
\runlabel{\(N_{\rm interval}=1\)} &
\xzpanel{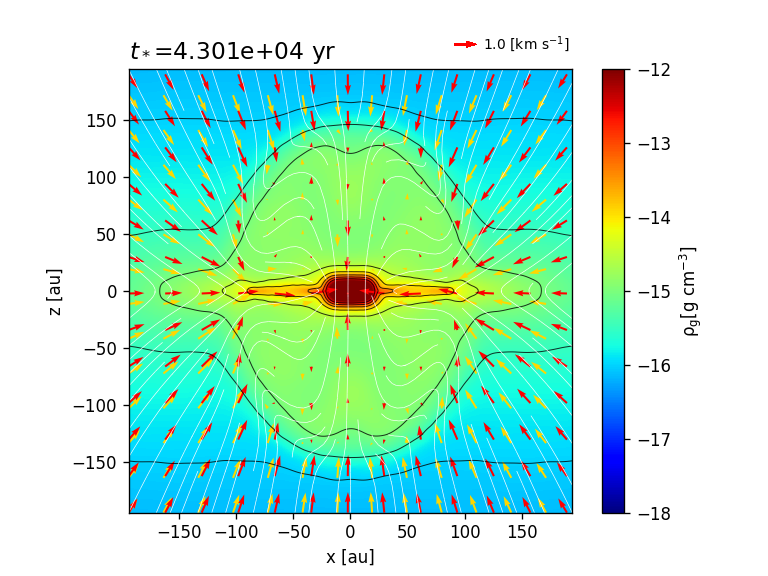} &
\xzpanel{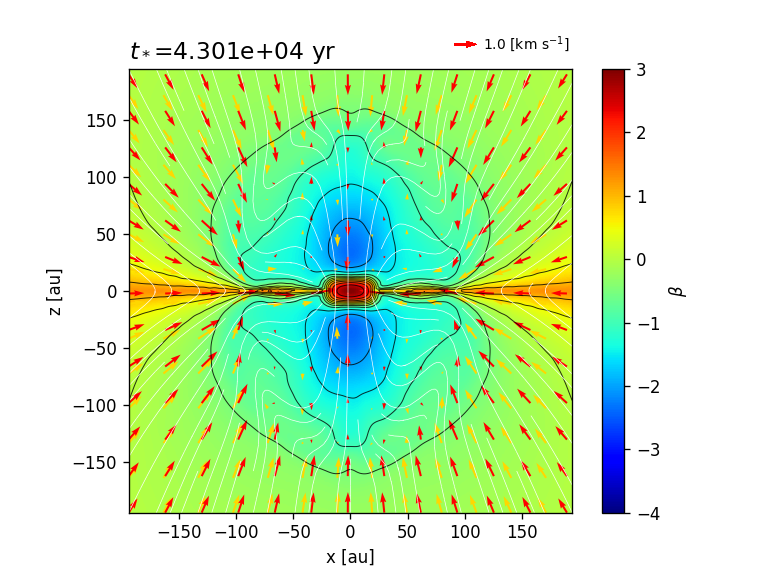} \\[-0.4ex]
\runlabel{\(N_{\rm interval}=10\)} &
\xzpanel{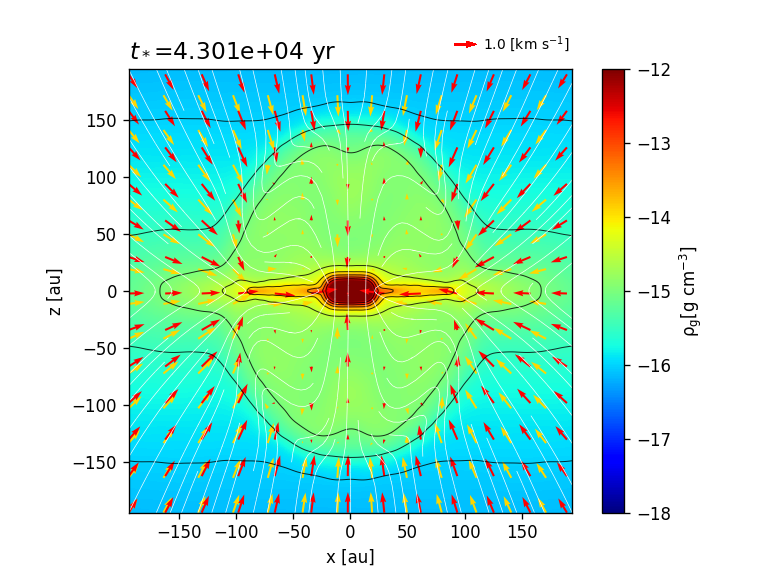} &
\xzpanel{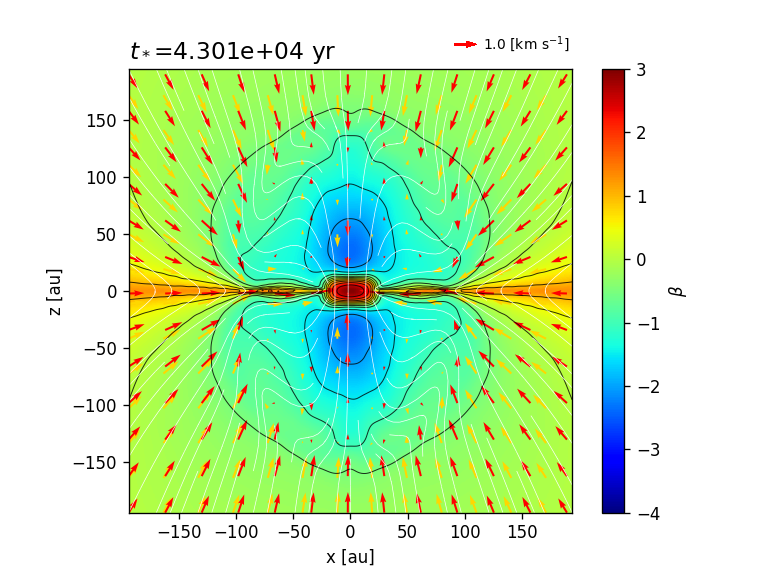} \\[-0.4ex]
\runlabel{\(N_{\rm interval}=100\)} &
\xzpanel{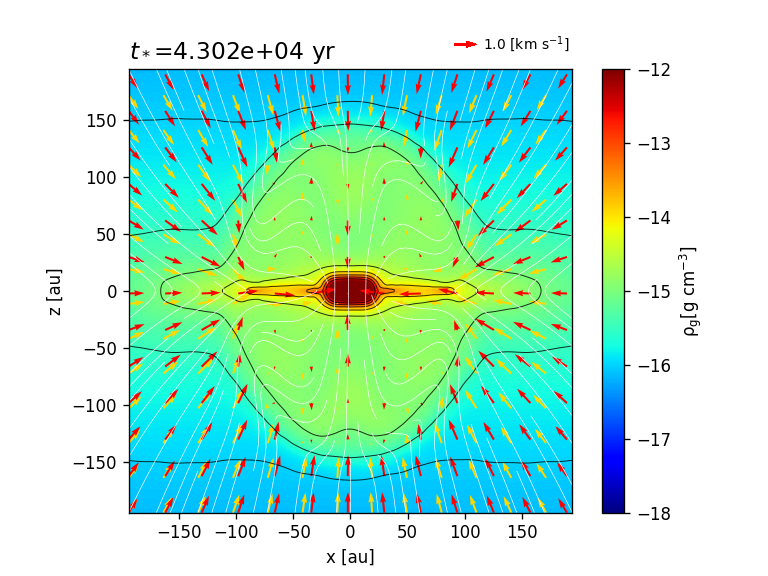} &
\xzpanel{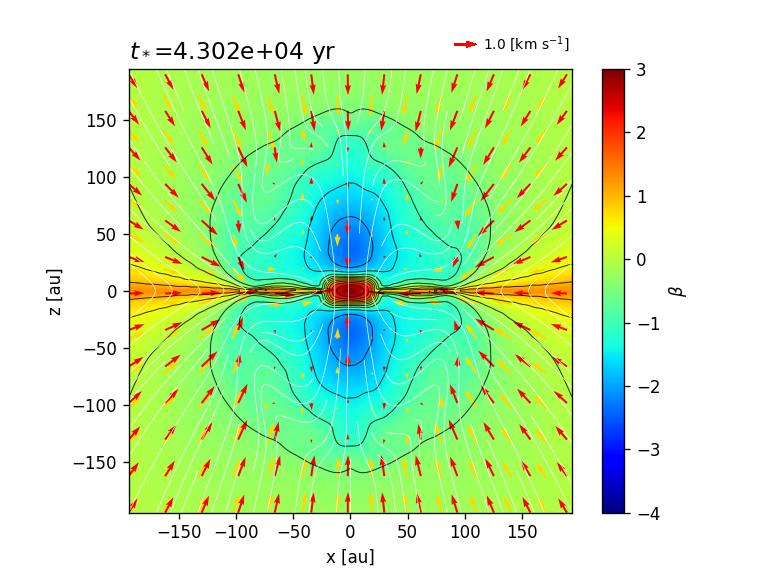} \\[-0.4ex]
\end{tabular}
}
\caption{Density and plasma \(\beta\) maps in the \(x\)-\(z\) plane of the gravitational collapse calculations at \(t\simeq4.3\times10^4\,{\rm yr}\).
The rows compare the divergence-cleaning run with projection runs using \(N_{\rm interval}=1\), 10, and 100.
The columns show density and plasma \(\beta\).}
\label{fig:collapse_xz_slices}
\end{figure*}

Figure \ref{fig:collapse_xz_slices} shows the \(x\)-\(z\) maps of the density and plasma \(\beta\) at \(t\simeq4.3\times10^4\,{\rm yr}\).
Although the time-averaged divergence error over a projection interval differs by roughly an order of magnitude between the \(N_{\rm interval}=1\), 10, and 100 runs, the density and plasma-\(\beta\) structures in the three projection runs are quantitatively very similar.
By contrast, the divergence-cleaning run is qualitatively similar to the projection runs, but quantitative differences are visible in the density structure and plasma \(\beta\) of the outflow just above the first core.
The plasma \(\beta\) inside the first core is also slightly lower in the divergence-cleaning run.
Similar quantitative differences between constrained-transport and divergence-cleaning calculations, as well as the dependence of the solution on the cleaning parameters, have been reported by \citet{2026arXiv260507928T}.
The consistency of the projection results over different values of \(N_{\rm interval}\) is therefore an important advantage of the present method.

\section{Summary and Discussion}

In this paper, we have studied a projection method for removing numerical \(\nabla\cdot\B\) errors in SPMHD.
In contrast to previous projection approaches, we constructed a gradient operator \(\G\) that is adjoint to the target discrete divergence operator \(D\).
We wrote both operators explicitly so that their linearity as discrete maps is transparent.
With this construction, we showed that the projection operator \(L=D\G=D M_V^{-1}D^T\) is symmetric positive semidefinite.
Consequently, the projection equation can be solved by the conjugate-gradient method.
We also showed that the projection is orthogonal with respect to the volume-weighted metric.
As a result, the correction does not increase the discrete magnetic energy, which is a desirable property for MHD calculations.

Using this method, we performed test calculations and showed that, with sufficiently many iterations, the divergence error can be reduced to the floating-point roundoff level.
One may worry that such roundoff-level projections are too expensive for practical simulations.
However, production calculations do not require such a strict tolerance.
It is sufficient to stop the projection once the divergence error has been reduced to the level generated by the underlying MHD update itself.
Our gravitational collapse calculations showed that this practical level of divergence control can be achieved at an affordable computational cost.
Even with this practical stopping criterion, the projection method produced substantially better results than the divergence-cleaning method that is commonly used in SPMHD.

The present method still contains free parameters that control when the projection is terminated and how often it is applied.
The need to choose these parameters is a weakness compared with constrained-transport schemes.
On the other hand, the parameters used here are dimensionless, and within the range examined in this paper, the simulation results are not strongly sensitive to parameters such as the projection interval.
This is an advantage of the present projection scheme over divergence cleaning, especially in view of the recent comparison by \citet{2026arXiv260507928T}, who showed that divergence-cleaning calculations can depend on the adopted wave speed and damping timescale.

Although we have focused on SPMHD, the same operator-based projection idea should also be applicable, with appropriate modifications, to other meshless or moving-mesh MHD schemes such as GIZMO \citep{2016MNRAS.455...51H} and AREPO.
We therefore consider this projection scheme to be an attractive alternative to divergence cleaning for controlling numerical \(\nabla\cdot\B\) errors in particle and meshless MHD calculations.

\section*{Acknowledgments}
The author thanks K. Tomida and S. Takasao for valuable discussions.
This work was supported by the JST FOREST Program, Grant Number JPMJFR2234.
The simulation code and data will be provided upon request from readers.

\bibliographystyle{elsarticle-num-names}
\bibliography{references,projection_extra}

@ARTICLE{2016JCoPh.322..326T,
       author = {{Tricco}, Terrence S. and {Price}, Daniel J. and {Bate}, Matthew R.},
        title = "{Constrained hyperbolic divergence cleaning in smoothed particle magnetohydrodynamics with variable cleaning speeds}",
      journal = {Journal of Computational Physics},
         year = 2016,
archivePrefix = {arXiv},
       eprint = {1607.02394},
 primaryClass = {astro-ph.IM},
          doi = {10.1016/j.jcp.2016.06.053}
}

@ARTICLE{2023PASJ...75..835T,
       author = {{Tsukamoto}, Yusuke and {Machida}, Masahiro N. and {Inutsuka}, Shu-ichiro},
        title = "{Co-evolution of dust grains and protoplanetary disks}",
      journal = {\pasj},
         year = 2023,
        month = oct,
       volume = {75},
       number = {5},
        pages = {835-852},
          doi = {10.1093/pasj/psad040},
archivePrefix = {arXiv},
       eprint = {2303.10419},
 primaryClass = {astro-ph.SR},
       adsurl = {https://ui.adsabs.harvard.edu/abs/2023PASJ...75..835T}
}

@ARTICLE{2011MNRAS.418.1668I,
   author = {{Iwasaki}, K. and {Inutsuka}, S.},
    title = "{Smoothed particle magnetohydrodynamics with a Riemann solver and the method of characteristics}",
  journal = {\mnras},
archivePrefix = "arXiv",
   eprint = {1106.3389},
 primaryClass = "astro-ph.GA",
     year = 2011,
    month = dec,
   volume = 418,
    pages = {1668-1688},
      doi = {10.1111/j.1365-2966.2011.19588.x},
   adsurl = {http://ads.nao.ac.jp/abs/2011MNRAS.418.1668I}
}

@ARTICLE{2012JCoPh.231.7214T,
   author = {{Tricco}, T.~S. and {Price}, D.~J.},
    title = "{Constrained hyperbolic divergence cleaning for smoothed particle magnetohydrodynamics}",
  journal = {Journal of Computational Physics},
archivePrefix = "arXiv",
   eprint = {1206.6159},
 primaryClass = "astro-ph.IM",
     year = 2012,
    month = aug,
   volume = 231,
    pages = {7214-7236},
      doi = {10.1016/j.jcp.2012.06.039},
   adsurl = {http://ads.nao.ac.jp/abs/2012JCoPh.231.7214T}
}

@ARTICLE{2005MNRAS.364..384P,
   author = {{Price}, D.~J. and {Monaghan}, J.~J.},
    title = "{Smoothed Particle Magnetohydrodynamics - III. Multidimensional tests and the {$\nabla\cdot\mathbf{B}=0$} constraint}",
  journal = {\mnras},
   eprint = {arXiv:astro-ph/0509083},
     year = 2005,
    month = dec,
   volume = 364,
    pages = {384-406},
      doi = {10.1111/j.1365-2966.2005.09576.x},
   adsurl = {http://ads.nao.ac.jp/abs/2005MNRAS.364..384P}
}

@ARTICLE{2002JCoPh.175..645D,
   author = {{Dedner}, A. and {Kemm}, F. and {Kr{\"o}ner}, D. and {Munz}, C.-D. and 
	{Schnitzer}, T. and {Wesenberg}, M.},
    title = "{Hyperbolic Divergence Cleaning for the MHD Equations}",
  journal = {Journal of Computational Physics},
     year = 2002,
    month = jan,
   volume = 175,
    pages = {645-673},
      doi = {10.1006/jcph.2001.6961},
   adsurl = {http://ads.nao.ac.jp/abs/2002JCoPh.175..645D}
}

@INPROCEEDINGS{2013ASPC..474..239I,
   author = {{Iwasaki}, K. and {Inutsuka}, S.},
    title = "{Hyperbolic Divergence Cleaning Method for Godunov Smoothed Particle Magnetohydrodynamics}",
booktitle = {Numerical Modeling of Space Plasma Flows (ASTRONUM2012)},
     year = 2013,
   series = {Astronomical Society of the Pacific Conference Series},
   volume = 474,
   editor = {{Pogorelov}, N.~V. and {Audit}, E. and {Zank}, G.~P.},
    month = apr,
    pages = {239},
   adsurl = {http://adsabs.harvard.edu/abs/2013ASPC..474..239I}
}

@ARTICLE{2018PASA...35...31P,
       author = {{Price}, Daniel J. and {Wurster}, James and {Tricco}, Terrence S. and {Nixon}, Chris and {Toupin}, Stephane and {Pettitt}, Alex and {Chan}, Conrad and {Mentiplay}, Daniel and {Laibe}, Guillaume and {Glover}, Simon and {Dobbs}, Clare and {Nealon}, Rebecca and {Liptai}, David and {Worpel}, Hauke and {Bonnerot}, Clement and {Dipierro}, Giovanni and {Ballabio}, Giulia and {Ragusa}, Enrico and {Federrath}, Christoph and {Iaconi}, Roberto and {Reichardt}, Thomas and {Forgan}, Duncan and {Hutchison}, Mark and {Constantino}, Thomas and {Ayliffe}, Ben and {Hirsh}, Kristian and {Lodato}, Giuseppe},
        title = "{Phantom: A Smoothed Particle Hydrodynamics and Magnetohydrodynamics Code for Astrophysics}",
      journal = {PASA},
         year = 2018,
        month = sep,
       volume = {35},
          eid = {e031},
        pages = {e031},
          doi = {10.1017/pasa.2018.25},
archivePrefix = {arXiv},
       eprint = {1702.03930},
 primaryClass = {astro-ph.IM},
       adsurl = {https://ui.adsabs.harvard.edu/abs/2018PASA...35...31P}
}

@ARTICLE{2000JCoPh.161..605T,
       author = {{Toth}, Gabor},
        title = "{The divergence constraint in shock-capturing magnetohydrodynamics codes}",
      journal = {Journal of Computational Physics},
         year = 2000,
        month = jul,
       volume = {161},
       number = {2},
        pages = {605-652},
          doi = {10.1006/jcph.2000.6519},
       adsurl = {https://ui.adsabs.harvard.edu/abs/2000JCoPh.161..605T}
}

@ARTICLE{1988ApJ...332..659E,
   author = {{Evans}, C.~R. and {Hawley}, J.~F.},
    title = "{Simulation of magnetohydrodynamic flows: A constrained transport method}",
  journal = {\apj},
     year = 1988,
    month = sep,
   volume = 332,
    pages = {659-677},
      doi = {10.1086/166684},
   adsurl = {https://ui.adsabs.harvard.edu/abs/1988ApJ...332..659E}
}

@ARTICLE{2012JCoPh.231..759P,
   author = {{Price}, D.~J.},
    title = "{Smoothed particle hydrodynamics and magnetohydrodynamics}",
  journal = {Journal of Computational Physics},
archivePrefix = "arXiv",
   eprint = {1012.1885},
 primaryClass = "astro-ph.IM",
     year = 2012,
    month = feb,
   volume = 231,
   number = 3,
    pages = {759-794},
      doi = {10.1016/j.jcp.2010.12.011},
   adsurl = {https://ui.adsabs.harvard.edu/abs/2012JCoPh.231..759P}
}

@ARTICLE{2016MNRAS.455...51H,
   author = {{Hopkins}, P.~F. and {Raives}, M.~J.},
    title = "{Accurate, meshless methods for magnetohydrodynamics}",
  journal = {\mnras},
archivePrefix = "arXiv",
   eprint = {1505.02783},
 primaryClass = "astro-ph.IM",
     year = 2016,
    month = jan,
   volume = 455,
   number = 1,
    pages = {51-88},
      doi = {10.1093/mnras/stv2180},
   adsurl = {https://ui.adsabs.harvard.edu/abs/2016MNRAS.455...51H}
}

@ARTICLE{2026arXiv260507928T,
   author = {{Tomida}, Kengo and {Sadanari}, Kenji Eric and {Takasao}, Shinsuke and
             {Iwasaki}, Kazunari},
    title = "{Systematic Comparison between Constrained Transport and Mixed Divergence Cleaning Methods for Astrophysical Magnetohydrodynamic Simulations}",
  journal = {arXiv e-prints},
archivePrefix = {arXiv},
   eprint = {2605.07928},
 primaryClass = {astro-ph.IM},
     year = 2026,
    month = may,
      eid = {arXiv:2605.07928},
    pages = {arXiv:2605.07928},
   adsurl = {https://ui.adsabs.harvard.edu/abs/2026arXiv260507928T}
}

\end{document}